\documentclass[aps,prd,preprint,superscriptaddress,nofootinbib]{revtex4-1}
\usepackage[english]{babel}
\usepackage[utf8]{inputenc}
\usepackage{amsmath,amssymb,braket,slashed,mathrsfs}
\usepackage{pifont}
\usepackage{graphicx}
\graphicspath{{./figures/}}
\usepackage{tikz}
\usetikzlibrary{shapes,arrows,positioning}
\tikzset{decision/.style={diamond, draw, fill=blue!20, text width=4.5em, text badly centered, inner sep=0pt}}
\tikzset{block/.style={rectangle, draw, fill=blue!20, text width=10em, text centered, rounded corners, minimum width=3.5cm}}
\tikzset{block1/.style={rectangle, draw, fill=blue!20, text width=18.5em, text centered, rounded corners, minimum width=3.5cm}}
\tikzset{line/.style={draw, -latex, thick}}
\usepackage{color,hyperref}
\usepackage{multirow,array}

\usepackage{cleveref}
\newcommand{\ba}{\begin{eqnarray}}
\newcommand{\ea}{\end{eqnarray}}

\newcommand{\nn}{\nonumber}
\newcommand{\MS}{\overline{\mathrm{MS}}}
\newcommand{\lat}{\mathrm{lat}}
\newcommand{\latt}{\mathrm{lat}}
\newcommand{\muRI}{\mu_0}
\newcommand{\RI}{\mathrm{RI}}

\definecolor{linkcolor}{rgb}{.17578125,.1875,.5703125}
\hypersetup{
    pdfstartview={FitH},    
    pdftitle={Rare kaon decays: a lattice study},    
    pdfauthor={RBC and UKQCD collaborations},           
    pdfsubject={Physics article},                
    pdfcreator={Xu Feng},          
    pdfkeywords={particle} {physics} {quantum} {chromodynamics} {QCD}  {hadrons} {kaons} {rare decays} {flavor} {FCNC} {lattice}, 
    colorlinks=true,        
    linkcolor=linkcolor,    
    citecolor=linkcolor,    
    filecolor=black,        
    urlcolor=linkcolor
}

\newcommand{\cu}{Physics Department, Columbia University, New York, NY 10027, USA}
\newcommand{\soton}{School of Physics and Astronomy, University of Southampton, 
Southampton SO17 1BJ, UK}
\newcommand{\uoe}{School of Physics and Astronomy, University of Edinburgh, Edinburgh EH9 3JZ, UK}

\begin{document}
\title{Prospects for a lattice computation of rare kaon decay amplitudes II $K\to\pi\nu\bar{\nu}$ decays}

\author{Norman~H.~Christ}\affiliation{\cu}
\author{Xu~Feng}\affiliation{\cu}
\author{Antonin~Portelli}\affiliation{\soton}\affiliation{\uoe}
\author{Christopher~T.~Sachrajda}\affiliation{\soton}
\collaboration{RBC and UKQCD collaborations}
\pacs{PACS}

\date{\today}

\begin{abstract}
The rare kaon decays $K\to\pi\nu\bar{\nu}$ are strongly suppressed in the standard model and widely regarded as processes in which new phenomena, not predicted by the standard model, may be observed.  Recognizing such new phenomena requires precise standard model prediction for the braching ratio of $K\to\pi\nu\bar{\nu}$ with controlled uncertainty for both short-distance and long-distance contributions.  In this work we demonstrate the feasibility of lattice QCD calculation of the long-distance contribution to rare kaon decays with the emphasis on $K^+\to\pi^+\nu\bar{\nu}$.  Our methodology covers the calculation of both $W$-$W$ and $Z$-exchange diagrams.  We discuss the estimation of the power-law, finite-volume    corrections and two methods to consistently combine the long distance contribution determined by the lattice methods outlined here with the short distance parts that can be reliably determined using perturbation theory.  It is a subsequent work of our first methodology paper on $K\to\pi\ell^+\ell^-$, where the focus was made on the $\gamma$-exchange diagrams. 
\end{abstract}

\maketitle

\section{Introduction}
    The ultra-rare kaon decays $K\to\pi\nu\bar{\nu}$ have attracted increasing interest 
    in recent decades.
    As flavor changing neutral current (FCNC) processes, these decays are highly 
    suppressed in the standard model (SM) and thus provide ideal probes for 
    the observation of new physics effects.
    In addition, the dominant, standard model contribution from the top quark
    loop to $K\to\pi\nu\bar\nu$  decays makes these processes very sensitive
    to the Cabibbo-Kobayashi-Maskawa (CKM) quark mixing matrix elements, $V_{ts}$ and $V_{td}$
    Therefore these decays
    can be used to determine $V_{td}$ in particular in a complementary and 
    independent manner to $B$ decays.

    Experimentally $K\to\pi \nu\bar{\nu}$ decays represent a 
    very substantial challenge.
    The first upper limit on the $K^+\to\pi^+\nu\bar{\nu}$ branching
    ratio was set by the heavy-liquid bubble chamber experiment in
    1969\,\cite{Camerini:1969hu}. It then took almost 30 years to actually observe the first
    $K^+\to\pi^+\nu\bar{\nu}$ event in the E787 experiment at the Brookhaven National Laboratory (BNL) in 1997\,\cite{Adler:1997am}.
    The current value
    for the branching ratio~\cite{Artamonov:2008qb} 
    \begin{equation}
    \textmd{Br}(K^+\to\pi^+\nu\bar{\nu})_{\textmd{exp}}=1.73^{+1.15}_{-1.05}\times10^{-10}
    \end{equation}
    is a combined result based on the 7 events collected by BNL
    E787~\cite{Adler:1997am,Adler:2000by,Adler:2001xv,Adler:2002hy} and
    its successor E949~\cite{Anisimovsky:2004hr,Artamonov:2008qb}. 
    The new experiment, NA62 at CERN~\cite{fortheNA62:2013jsa}, 
    aims at an observation of $O(100)$ events and a 10\%-precision measurement of
    ${\rm Br}(K^+\to\pi^+\nu\bar{\nu})$.
    In the coming decades $K^+\to\pi^+\nu\bar{\nu}$ decays are therefore likely to lead to  
    precision determinations of the SM parameters 
    and stringent tests of possible effects of new physics.
     
    The search for the decays  $K_L\to\pi^0\nu\bar{\nu}$, with only neutral particles in the initial and final states, is
    even more challenging experimentally. Indeed,
    $K_L\to\pi^0\nu\bar{\nu}$ events have never been observed and currently there is only the 
    upper bound for the  branching ratio 
    \begin{equation}\label{eq:KLexpbound}
    {\rm Br}(K_L\to\pi^0\nu\bar{\nu})\le
    2.6\times10^{-8}\quad\textrm{at~90\%~confidence level\,,}\end{equation}
    set by the E391a experiment at the 12\,GeV proton synchrotron at KEK in 2010\,\cite{Ahn:2009gb}. 
    This bound is three orders of magnitude larger than a recent SM prediction~\cite{Buras:2015qea}\, 
    \begin{equation}
    {\rm Br}(K_L\to\pi^0\nu\bar{\nu})_{\textrm{SM}}=\left(3.00\pm0.30\right)\times 10^{-11}\,.\end{equation}
    The new KOTO experiment at J-PARC~\cite{Iwai:2012qya} will be sensitive to much lower branching ratios than that given by the bound in Eq.\,(\ref{eq:KLexpbound}), indeed to ones also below the Grossman-Nir model-independent upper bound\,\cite{Grossman:1997sk},
     $\textrm{Br}(K_L\to\pi^0\nu\bar{\nu})<4.4\,\textmd{Br}(K^+\to\pi^+\nu\bar{\nu})$\,.  KOTO will thus explore much of the parameter space of theories beyond the standard model (BSM).
    
    On the theoretical side, $K\to\pi\nu\bar{\nu}$ decays
    are known to be short-distance (SD) dominated. The required hadronic matrix elements 
    can be obtained from measurements of charged-current semi-leptonic kaon decays, such as $K^+\to\pi^0e^+\nu$ decays. We will explain in more detail in the next section that the long-distance (LD) contributions, i.e. contributions from distances on the order of, or larger than, the inverse of the mass of the charm quark,
     are safely neglected in $K_L\to\pi^0\nu\bar{\nu}$ decays and
    are expected to be small in $K^+\to\pi^+\nu\bar{\nu}$ decays.
However, a lattice QCD calculation of these effects may be required to convincingly establish their size and will become necessary when a precise comparison between the SM prediction and the NA62 or future measurements is required. The purpose of this paper is to set out the framework necessary for the lattice computation of long-distance effects in $K^+\to\pi^+ \nu\bar{\nu}$ decays.

    In our earlier paper~\cite{Christ:2015aha} we had proposed a method for the computation of $K\to\pi\ell^+\ell^-$ decay amplitudes (where $\ell$ is a charged lepton) using lattice QCD and focussing on the dominant $\gamma$-exchange diagrams.
    In this work we extend the discussion to $K\to\pi\nu\bar{\nu}$ decays which requires us to include the $W$-$W$
    and $Z$-exchange diagrams. In addition to Ref.\,\cite{Christ:2015aha}, our work builds on several other earlier studies.
    In Ref.~\cite{Isidori:2005tv} it had been first proposed to use lattice QCD to calculate the LD contributions to rare kaon decay amplitudes, including those for 
    $K\to\pi\nu\bar{\nu}$ decays. That paper focussed on the ultraviolet divergences which appear in the integral over the separation of the two operators  (two weak operators in the case of $K\to\pi\nu\bar{\nu}$ decays) as the two operators approach each other. For the $\gamma$-exchange diagrams which give the dominant contribution to  $K\to\pi\ell^+\ell^-$ decays, the authors stressed the importance of using the conserved electromagnetic current to reduce the degree of divergence and to control this short-distance divergence. For the axial current, necessarily present when calculating $K\to\pi\nu\bar\nu$ decay amplitudes, this is a more involved problem, particularly with the use of Wilson fermions considered in Ref.~\cite{Isidori:2005tv}. Below we explain how to deal with the corresponding SD divergences when using domain wall fermions, a formulation which respects chiral symmetry to good precision.
We have also benefited from the methods developed by the RBC-UKQCD collaboration in their computations of long-distance effects in second-order electroweak processes\,\cite{Christ:2010gi,Christ:2012np}; methods which have been successfully applied to 
    the lattice calculation of the $K_L$-$K_S$ mass difference~\cite{Christ:2012se,Bai:2014cva} and are currently being applied to the evaluation of the long-distance contribution to the indirect CP-violating parameter $\epsilon_K$~\cite{Bai:2015xxx}.
    
The paper is organized as follows: We first introduce the phenomenological background for $K\to\pi\nu\bar{\nu}$ decay with an emphasis on the LD contributions in Section~\ref{sec:phenkpinunu}. Then, in Section~\ref{sec:methodology}, we describe the detailed methodology proposed to calculate this long-distance part using lattice QCD, specifically for the case of $K^+\to\pi^+\nu\bar{\nu}$.  The technical issue of how to use the standard, perturbative, short-distance result for $K^+\to\pi^+\nu\bar{\nu}$ to determine the new low energy constant that appears in the second-order effective theory used in our lattice calculation is described in Section~\ref{sec:SD}.  In Section~\ref{sec:FV} we discuss the power-law, finite-volume effects which must be subtracted in order to obtain the physical, infinite volume result with sufficient precision.  A summary and conclusions are presented in Section~\ref{sec:conclusion}.  Finally Appendixes~\ref{sec:M-E}, \ref{sec:states}, \ref{sec:WW_scalar_amp} and \ref{sec:ground_state_WW} describe the relation between the Minkowski- and Euclidean-space ampliutdes used in this paper, the conventions adopted for the mesonic and lepontic states, the extraction of the scalar amplitude $F_{WW}(p_K,p_\nu,p_{\bar{\nu}})$ characterizing the $W$-$W$ exchange diagrams and the method used to remove the unphysical contribution of intermediate states with energy below $M_K$, respectively.

\section{Phenomenological background}
\label{sec:phenkpinunu}
In the SM $K\to\pi\nu\bar{\nu}$ decays are second-order electroweak processes, involving
$W$-$W$ exchange diagrams (diagrams which contain two $W$-boson exchanges) and $Z$-exchange diagrams (diagrams
which contain a $W$- and $Z$-boson or a $W$-$W$-$Z$ vertex).
    As explained below, the dominant contribution comes from diagrams in which a top quark propagator explicitly appears.  The corresponding contribution from the propagation of the charm quark is suppressed by a factor of
    $(m_c/M_W)^2$ through the Glashow-Iliopoulos-Maiani (GIM) mechanism
    but is enhanced by a factor of $\log M_W/m_c$. Here $m_c$ and $M_W$ are the masses of the charm quark and $W$-boson respectively.
    In the CP-violating decay $K_L\to\pi^0\nu\bar{\nu}$, the amplitude
    depends on the imaginary parts of the CKM matrix elements and this provides a further suppression of the charm-quark contribution.   As a result of the strong suppression of the charm quark contribution, this decay is completely SD dominated and is one of the theoretically
    cleanest places to search for the effects of new physics. The absence of LD contributions implies that a lattice QCD calculation of $K_L\to\pi^0\nu\bar{\nu}$ decays is
    unnecessary.

    The situation is different however, for the CP-conserving decays 
    $K_S\to\pi^0\nu\bar{\nu}$ and $K^+\to\pi^+\nu\bar{\nu}$. For these decays
    the real parts of the CKM matrix elements enhance the charm quark
    contribution (estimated to be about $\sim29\%$ of the total amplitude\,\cite{Cirigliano:2011ny}) and even the contribution of the up quark
    is not completely negligible ($\sim3\%$ of the total amplitude\,\cite{Cirigliano:2011ny}). 
    
    The decay length of the $K_S$ meson is so short that $K_S\to\pi^0\nu\bar\nu$ decays are currently unobservable experimentally. The CERN NA62 experiment, with its higher energy beam, could in principle place the detector close enough to the target but studies are still required to see whether it could withstand the high intensities which would be present~\footnote{A.~Cecucci and C.~Lazzeroni, private communication}. KOTO instead has a low energy beam which results in a decay length which is too short to be observed. We therefore concentrate our investigation on the $K^+\to\pi^+\nu\bar{\nu}$ decays which are already being studied by the NA62 experiment, with data taking having started in the summer of 2015~\cite{fortheNA62:2013jsa}.

    In contrast to the $K_L-K_S$ mass difference, where the charm quark contribution has a large non-perturbative component~\cite{Brod:2011ty,Bai:2014cva,Christ:2012se}, for   $K^+\to\pi^+\nu\bar{\nu}$ decays the contribution of the charm quark is expected to be predominantly perturbative and come from SD effects.  A one-loop perturbative calculation of
the electroweak interactions performed by Inami and Lim~\cite{Inami:1980fz} shows that  the charm quark contribution to the decay amplitude is proportional to $-\frac{3}{4}x_c\log x_c-\frac{1}{4}x_c$, where $x_c=m_c^2/M_W^2$.  Here, the logarithmic term $x_c\log x_c$ is the largest part of the charm contribution, which suggests that the dominant energy scale lies between $M_W$ and $m_c$.  However, when the leading-log QCD corrections, which sum those terms of the form $x_c\alpha_s^n\ln^{n+1}x_c$ to all orders in $\alpha_s$, are included it is found that the SD, charm-quark contribution is suppressed by 35\%~\cite{Ellis:1982ve,Dib:1989cc,Buchalla:1990qz}, relative to the leading-order, Inami-Lim result. This large suppression has two consequences.  First it motivates the work to include the SD QCD effects to higher orders in 
perturbation thoery~\cite{Buchalla:1993wq,Buras:2005gr,Buras:2006gb}.  Second it gives  increased importance to the LD QCD contributions coming from energy scales at or below the charm quark mass.  This makes the first-principles, lattice calculation of these LD QCD effects increasingly necessary for the comparison between SM predictions and future experimental results for this decay.

    A very recent SM prediction for the $K^+\to\pi^+\nu\bar{\nu}$ branching ratio is
    given by~\cite{Buras:2015qea}
    \begin{equation}
    \label{eq:SM_rarek_neutrino}
    \textmd{Br}(K^+\to\pi^+\nu\bar{\nu})_{\textmd{SM}}=
    \left(9.11\pm 0.72\right)
    \times 10^{-11}\,.
    \end{equation}
    To understand the origin of the uncertainty in Eq.\,(\ref{eq:SM_rarek_neutrino}), we write the branching ratio as in Eq.\,(4.5) of Ref.\,\cite{Brod:2010hi}:
    \begin{equation}\label{eq:Brtheory}
    \textmd{Br}(K^+\to\pi^+\nu\bar{\nu})_{\textmd{SM}}=
    \kappa_+(1+\Delta_{\mathrm{EM}})\cdot
    \left[\left(\frac{\mathrm{Im}\lambda_t}{\lambda^5}X_t(x_t)\right)^{\!\!2}
    +\left(\frac{\mathrm{Re}\lambda_c}{\lambda}P_c+\frac{\mathrm{Re}\lambda_t}{\lambda^5}X_t(x_t)\right)^{\!\!2}\,\right]\,.
    \end{equation}
    In Eq.\,(\ref{eq:Brtheory}), $\Delta_{\mathrm{EM}}$ is the electromagnetic correction, $\lambda=|V_{us}|$ 
    and $\lambda_q=V_{qs}^*V_{qd}$ are CKM (or products of CKM) matrix elements,
    $X_t(x_t)$ is the top-quark contribution (with $x_t=m_t^2/M_W^2$) and $P_c$ is 
    the total charm quark contribution. More precisely, we have included the up quark contribution in 
    both $X_t$ and $P_c$, eliminating $\lambda_u$ by using the unitarity relation $\lambda_u+\lambda_c+\lambda_t=0$. 
    We distinguish two contributions to $P_c$ 
    \begin{equation}\label{eq:Pc}
    P_c=P_c^{\mathrm{SD}}+\delta P_{c,u}\,,
    \end{equation}
    where $P_c^{\mathrm{SD}}$ is the SD contribution coming from energy scales above the charm quark mass. 
    The remaining LD contribution, denoted as $\delta P_{c,u}$, includes contributions from both the charm and 
    up quark loops. 
    The parameter $\kappa_+$ in Eq.\,(\ref{eq:Brtheory}) contains the remaining factors,
    including the hadronic matrix element from semi-leptonic $K^+$ decay.

    The dominant uncertainty in Eq.~\eqref{eq:SM_rarek_neutrino} arises from the SM input parameters, 
    especially the CKM matrix elements.
    Because of the dominance of the top quark contribution $X_t(x_t)$,
    the CKM matrix elements in $\lambda_t$ associated with the top quark have a
    large impact on the branching ratio. In order to make a more precise SM prediction it is
    therefore necessary to know these CKM matrix elements more accurately.
    On the other hand, as a result of higher-order perturbative calculations,
    especially the NLO QCD~\cite{Buchalla:1998ba,Misiak:1999yg} and the two-loop electroweak corrections~\cite{Brod:2010hi} to the top quark contribution $X_t(x_t)$, as well as the NNLO QCD~\cite{Buras:2005gr,Buras:2006gb} and the NLO electroweak  corrections~\cite{Brod:2008ss} to the charm quark contribution $P_c^{\mathrm{SD}}$, the omitted, higher-order perturbative effects in the top and SD charm quark contributions are no longer the main source of theoretical uncertainty.

    Although the size of the
    LD contribution is estimated to be
    small, it now contributes a significant, if still sub-dominant, source for the SM uncertainty.
    Ref.~\cite{Isidori:2005xm} gives a phenomenological estimate of this
    LD effect based on chiral perturbation theory and the operator
    production expansion.  The resulting estimate of the LD contribution,
    $\delta P_{c,u}=0.04\pm0.02$, enhances the branching ratio
    $\textmd{Br}(K^+\to\pi^+\nu\bar{\nu})_{\textmd{SM}}$ by 6\%, which is
    comparable to the 8\% total SM parametric error given in
    Eq.~(\ref{eq:SM_rarek_neutrino}).  Here the quoted $\pm0.02$ error is
    necessarily a rough estimate which cannot easily be systematically improved. This
    quoted error translates into a 3\% uncertainty for the branching ratio, but it  is possible that the LD contribution might be somewhat larger or even much smaller than this estimate. 
    We do not have a clear answer at present and this provides the motivation for the development of lattice techniques to compute these LD contributions.

    Lattice QCD can provide a first-principles determination of the LD
    contribution with controlled errors. 
    Therefore it was proposed in Ref.~\cite{Isidori:2005tv} and 
    endorsed in Ref.~\cite{Buras:2006gb}
    to perform a direct lattice QCD calculation of the LD contribution
    to $K^+\to\pi^+\nu\bar{\nu}$ decay amplitudes. 
    Recognizing that the SM predictions will be confronted with new NA62 measurements in the near future, 
    it is timely to have a lattice QCD calculation of these LD effects. 


\section{Method}
\label{sec:methodology}
    Since the dominant contribution to the $K^+\to\pi^+\nu\bar{\nu}$ amplitude comes from  
    the top quark loop and the sub-leading charm quark contribution
    is also SD dominated, it is natural to write these contributions in terms of the matrix element of a 
    low-energy effective Hamiltonian
    \begin{equation}\label{eq:A6}
    A_0(K^+\to\pi^+\nu\bar{\nu})=\langle\pi^+\nu\bar{\nu}|{\mathcal H}_{\textrm{eff},0}|K^+\rangle,
    \end{equation}
    where ${\mathcal H}_{\textrm{eff},0}$ is given in terms of the dimension-six local 
    operator $Q_0=(\bar{s}d)_{V-A}\,(\bar{\nu}_\ell \nu_\ell)_{V-A}$~\cite{Buchalla:1993wq,Buchalla:1998ba}:
    \begin{eqnarray}
    \label{eq:effHamiltonian}
    {\mathcal H}_{\textrm{eff},0}=\frac{G_F}{\sqrt{2}}\frac{\alpha}{2\pi\sin^2\theta_W}
    \sum_{\ell=e,\mu,\tau}\left[\lambda_t X_t(x_t)+\lambda_c X_c^\ell(x_c)\right]Q_0,
    \end{eqnarray}
    and  $x_q=m_q^2/M_W^2$.
    Here $G_F$ is the Fermi constant, $\alpha$ is the fine-structure constant and 
    $\theta_W$ is the Weinberg weak mixing angle. The Inami-Lim functions $X_t(x_t)$ and $X_c^\ell(x_c)$ 
    are the Wilson coefficients,
    representing the contributions of the internal top quark and charm quark to the
    operator $Q_0$. They were first calculated by Inami and Lim in 1980 at one-loop order~\cite{Inami:1980fz}.
    As in Section\,\ref{sec:phenkpinunu}, we eliminate $\lambda_u$ by using the unitarity relation
    $\lambda_u=-\lambda_c-\lambda_t$ and absorbing the contribution from the $u$-quark in $X_t$ and $X_c^\ell$, in which we set
    $x_u=0$.  In Eq.~(\ref{eq:effHamiltonian}) the top and charm quark
    degrees of freedom have both been integrated out. The remaining hadronic effects are contained in the matrix element
    $\langle\pi^+|(\bar{s}d)_{V-A}|K^+\rangle$, which, in the isospin-symmetric limit, is the same matrix element as that containing the non-perturbative QCD effects in $K_{\ell 3}$ decays.
    
    The $X_c^\ell$ in Eq.\,(\ref{eq:effHamiltonian}) are related to $P_c^{\textrm{SD}}$ in Eq.\,(\ref{eq:Pc}) by
    \begin{equation}
    P_c^{\textrm{\,SD}} = \frac1{\lambda^4} \, \frac{X_c^e(x_c)  + X_c^\mu(x_c) + X_c^\tau(x_c)}{3}\,,
    \end{equation}where the factor of 3 in the denominator performs the conventional average of $X_c^\ell$ over the three lepton flavours. The subscript $\ell$ on $X_t$ is not included since the lepton mass dependence is suppressed by a factor of $(m_\ell/m_t)^2$ which can be neglected even for the $\tau$-lepton. For the charm quark contribution the lepton mass dependence cannot be neglected, particularly for the $\tau$-lepton, and hence the superscript $\ell$ is introduced for this case. 
    
    The contribution $A_0(K^+\to\pi^+\nu\bar{\nu})$ in Eq.\,(\ref{eq:A6}), obtained using the local effective Hamiltonian ${\mathcal H}_{\textrm{eff},0}$ in Eq.\,(\ref{eq:effHamiltonian}), accurately reproduces the contribution from the top quark and 
the SD component of the charm quark contribution. Of course it does not contain the LD component of 
    the charm quark contribution which is intrinsically bilocal. The evaluation of this long distance contribution is the main subject of this paper and we now begin our discussion of this.
    
    To explore the bilocal structure of the up- and charm-quark contributions,
    we begin with the first-order effective field theory, where the $W$ and $Z$ bosons have been integrated out. The bilocal contributions are constructed from two insertions of the first-order effective Hamiltonian.
    The four-Fermi, effective weak Hamiltonian relevant for the $K^+\to\pi^+\nu\bar{\nu}$ decay amplitudes can be written as~\cite{Buchalla:1995vs,Isidori:2005tv}
    \begin{eqnarray}
    \label{eq:LO_Heff}
    {\mathcal H}_{\textrm{eff}}^{\textrm{LO}}=\frac{G_F}{\sqrt{2}}\sum_{q,\ell}\left(V_{qs}^*O_{q\ell}^{\Delta S=1}+V_{qd}O_{q\ell}^{\Delta S=0}\right)
    +\frac{G_F}{\sqrt{2}}\sum_{q}\lambda_q O_q^W+\frac{G_F}{\sqrt{2}}\sum_\ell O_\ell^Z\,,
    \end{eqnarray}    
    where the sums over the quarks $q$ run over $q=u,c$ and those over the leptons $\ell$ run over $\ell=e,\mu,\tau$. 
    
    The first term on the right hand side of Eq.~(\ref{eq:LO_Heff}) results from the $W$-$W$ diagrams, in which the $W$-boson exchanges have been
    replaced by two effective operators
    \begin{eqnarray}
   O_{q\ell}^{\Delta S=1}&=&C_{\Delta S=1}^{\MS}(\mu)\left[(\bar{s}q)_{V-A}\,(\bar{\nu}_\ell\ell)_{V-A}\right]^{\MS}(\mu),\nn\\
   O_{q\ell}^{\Delta S=0}&=&C_{\Delta S=0}^{\MS}(\mu)\left[(\bar{\ell}\nu_\ell)_{V-A}\,(\bar{q}d)_{V-A}\right]^{\MS}(\mu),\label{eq:WW_operator}
    \end{eqnarray}
where for fermion fields $f_{i}$, ($i=1$-4) 
\begin{equation}
(\bar{f}_1f_2)_{V-A}\,(\bar{f}_3f_4)_{V-A}\equiv(\bar{f}_1\gamma_\mu (1-\gamma^5)f_2)~(\bar{f}_3\gamma_\mu (1-\gamma^5)f_4)\,.
\end{equation}
   We absorb the Wilson coefficients $C_{\Delta S=1}^{\MS}(\mu)$ and $C_{\Delta S=0}^{\MS}(\mu)$ into the definition of the operators $O_{q\ell}^{\Delta S=1}$ and $O_{q\ell}^{\Delta S=0}$.  Here and below we will find it convenient to use the letter $O$ to represent an operator which incorporates a Wilson coefficient and the letter $Q$ for an operator which does not include such a coefficient.
   These coefficients account for the contributions from SD physics and are conventionally and conveniently calculated in the $\MS$ scheme.
   For the particular operators appearing in Eq.~\eqref{eq:WW_operator}, the Ward-Takahashi identity implies $C_{\Delta S=1}^{\MS}(\mu)=C_{\Delta S=0}^{\MS}(\mu)=1$.
   The quark current operators renormalized in the $\MS$ scheme can be related to the 
   bare lattice operators by $[(\bar{q}q')_{V/A}]^{\MS}=Z_{V/A}[(\bar{q}q')_{V/A}]^{\lat}$.
   Here $Z_V$ and $Z_A$ are the renormalization constants for vector and axial-vector currents.  They are quark-mass and renormalization scale independent up to lattice artifacts.  If the conserved lattice current operators are used in a (almost) chirally symmetric formulation of lattice QCD, such as domain wall fermions, then $Z_V=Z_A=1$.  For simplicity in the remainder of the paper we will neglect the $O(a^2)$ effects which distinguish $Z_A$ from $Z_V$ and replace $Z_A$ with $Z_V$, which will be assumed to be quark mass and scale independent.
   
   The second and third terms on the right-hand side of Eq.~(\ref{eq:LO_Heff}) 
   are relevant for the $Z$-exchange diagrams. Note that these diagrams include the 
exchanges of both a $W$- and $Z$-boson. The $W$-boson exchange is described by 
the four-quark operator $O_q^W$, 
   \begin{eqnarray}
   \label{eq:O_W_operator}
   O_q^W=C_1^{\MS}(\mu)\,Q_{1,q}^{\MS}(\mu)
   +C_2^{\MS}(\mu)\,Q_{2,q}^{\MS}(\mu),
   \end{eqnarray}
   where $Q_{i,q}^{\MS}(\mu)$ ($i=1,2$) are conventional current-current
   operators renormalized in the $\MS$ scheme. They can be related to the bare lattice operators by a matrix of renormalization constants $Z_{i,j}^{\lat\to\MS}(a\mu)$
   \begin{eqnarray}
   &&Q_{i,q}^{\MS}(\mu)=\sum_jZ_{i,j}^{\lat\to\MS}(a\mu)\,Q_{j,q}^{\lat}(a),\quad i,j=1,2, \quad\textrm{where}\label{eq:O12qMSbar}
   \\
   &&\hspace{-0.2in}Q_{1,q}^{\lat}=(\bar{s}_a q_b)_{V-A}\,(\bar{q}_b d_a)_{V-A},\quad
   Q_{2,q}^{\lat}=(\bar{s}_a q_a)_{V-A}\,(\bar{q}_b d_b)_{V-A}\label{eq:O12qlat}
   \end{eqnarray}
   and $a,b$ are color indices.
   The detailed procedure to compute the renormalization matrix $Z_{i,j}^{\lat\to\MS}(a\mu)$ 
   can be found in Refs.~\cite{Blum:2001xb,Blum:2011pu,Christ:2012se}.
   Note that the $\mu$-scale dependence in the Wilson coefficients $C_i^{\MS}(\mu)$
   and the renormalized operators $Q_{i,q}^{\MS}(\mu)$ cancels, leaving
   the operator $O_q^W$ scale independent.
   The exchange of the $Z$-boson propagator has been replaced by a two-quark-two-neutrino operator $O^Z_\ell$
   \begin{eqnarray}
   \label{eq:O_Z_operator}
   O^Z_\ell=C_Z^{\MS}(\mu)\left[J_\mu^Z\,\bar{\nu}_\ell\gamma^\mu(1-\gamma_5) \nu_\ell\right]^{\MS}(\mu)
   \end{eqnarray}
   where the neutral current $J_\mu^Z$ is given by 
   \begin{equation}\label{eq:JmuZ}
   J_\mu^Z=\sum_{q=u,c,d,s}(T_3^q\,\bar{q}\gamma_\mu(1-\gamma_5)q 
   -2Q_{\textrm{em},q}\sin^2\theta_W\,\bar{q}\gamma_\mu q).\end{equation} The weak isospin $T_3^q$ and the electric charge $Q_{\textrm{em},q}$ take
   the values $+\frac12$ and $+\frac23$ respectively for $q=u$ and $c$ and the values $-\frac12$ and $-\frac13$ for $q=d$ and $s$.
   As described above, we have $C_Z^{\MS}(\mu)=1$. The quark current operators renormalized
   in the $\MS$ scheme can be related to the bare lattice operator by 
   $[J_\mu^Z]^{\MS}=Z_V[J_\mu^Z]^{\lat}$.

   As the next step we work to second order in the standard, non-renormalizable,
   effective field theory of the weak interactions and construct the bilocal product of 
   two first-order, four-fermi effective operators from Eq.\,(\ref{eq:LO_Heff})
   as follows:
    \begin{equation}\label{eq:bilocal}
    {\mathcal B}(y)=\frac{G_F}{\sqrt{2}}\frac{\alpha}{2\pi\sin^2\theta_W}\frac{\pi^2}{M_W^2}\lambda_c\,
    \sum_{\ell=e,\mu,\tau}\Big({\mathcal B}_{WW}(y)+{\mathcal B}_{Z}(y)\Big)
    \end{equation}
    where
    \begin{equation}\label{eq:B_WW}
    {\mathcal B}_{WW}(y)=\int d^4x\,T[O_{u\ell}^{\Delta S=1}(x)O_{u\ell}^{\Delta S=0}(y)]
    -\{u\to c\}
    \end{equation}
    and
    \begin{equation}
     \label{eq:B_Z}
    {\mathcal B}_{Z}(y)=\int d^4x\,T[O_u^W(x)O_\ell^Z(y)]-\{u\to c\}.
    \end{equation}
    For compactness of notation we have suppressed the label $\ell$ in ${\mathcal B}_{WW}(y)$
    and ${\mathcal B}_{Z}(y)$, but the reader should note that there is such a dependence.  
    We should also point out that in Eq.~\eqref{eq:B_WW} we have made an arbitrary choice of which of the two operators is integrated over space-time and which is evaluated at the fixed position $y$.
    The bilocal product ${\mathcal B}(y)$ has been separated into two parts, ${\mathcal B}_{WW}(y)$
    and ${\mathcal B}_{Z}(y)$, the first associated with $W$-$W$ diagrams
    and the second with $Z$-exchange diagrams.
    The minus sign in Eqs.~(\ref{eq:B_WW}) and (\ref{eq:B_Z}) comes from the 
    GIM mechanism under the approximation of $\lambda_u\approx -\lambda_c$. Here the bilocal product
    ${\mathcal B}(y)$ is defined in Euclidean space to favor a lattice QCD calculation.
    Its Minkowski-space definition can be found in Ref.~\cite{Buchalla:1993wq}.

In infinite-volume calculations of matrix elements, performing an integral over $y$ in Eqs.~\eqref{eq:B_WW} and \eqref{eq:B_Z} would introduce a four-dimensional, momentum-conserving delta function.  In computations using lattice QCD, which are necessarily performed in a finite volume, this delta-functions is replaced by a factor of the space-time volume. As will be described in greater detail below, for the $K^+\to\pi^+\nu\overline{\nu}$ decay amplitude discussed in this paper we propose to integrate $y$ over the full spatial volume and to integrate the times at which each of the operators are evaluated over a fixed interval $[-T_a,T_b]$, chosen to lie sufficiently far from the initial kaon and final pion to suppress possible excited hadronic-state contamination.  This follows closely the procedure used earlier in the calculation of the $K_L - K_S$ mass difference~\cite{Christ:2012se}.  

  The second-order $K^+\to\pi^+\nu\bar\nu$ decay amplitude can be obtained 
   by evaluating matrix elements of the bilocal operators ${\mathcal B}_{WW}$ and ${\mathcal B}_{Z}$
   and a third (local) operator $C_0 Q_0^{\MS}$:
   \begin{eqnarray}\label{eq:ct}
  {\cal A}(K^+\to\pi^+\nu\bar\nu)\propto \langle\pi^+\nu\bar{\nu}|{\mathcal B}_{WW}(0)+{\mathcal B}_Z(0)|K^+\rangle +
   \langle\pi^+\nu\bar{\nu}|C_0 Q_0^{\MS}(0)|K^+\rangle\,,
   \end{eqnarray}
   where $C_0$ is a Wilson coefficient and
   $Q_0^{\MS} =(\bar{s}d)_{V-A}(\bar{\nu}\nu)_{V-A}^{\MS}$
   a local operator renormalized in the $\MS$ scheme. 
   Here $C_0 Q_0^{\MS}$ is a regulator-dependent counter term
   which removes the new ultra-violet singularities in ${\mathcal B}_{WW}$ and ${\mathcal B}_Z$ that arise when two of the 
   dimension-six, four-fermi operators which appear in the same
   diagram approach each other.  The need for such added counter terms is a standard
   feature of a non-renormalizable effective theory and is discussed at length in Sect.~\ref{sec:SD}.

   The presence of this $C_0 Q_0^{\MS}$ counter term reflects a new renormalization
   constant that must be introduced when the effective theory is
   evaluated at second order and that must be determined using some
   additional physical input. For the case of the weak interactions,
   this new renormalization constant $C_0$ must be determined
   by requiring that the effective theory, evaluated at second order
   agrees with the second-order predictions of the underlying SM. 
   A convenient way to formulate such a requirement is to
   impose ``Rome-Southampton'' conditions on the second-order 
   $\bar{s}d$\,-\,$\bar{\nu}\nu$ Green's function, which corresponds to the
   $K^+\to\pi^+\nu\bar\nu$ decay, demanding that this Green's
   function, evaluated at a momentum scale $\mu_0$, agrees when
   evaluated in both the effective theory and the SM.   
   If infra-red safe, non-exceptional momenta are chosen when
   applying the Rome-Southampton condition, as described in
   Sec.~\ref{sec:SD}, and the scale $\mu_0$
   is chosen much larger than the scale of QCD,
   $\mu_0 \gg \Lambda_{\mathrm{QCD}}$, then the required SM calculation 
   can be accurately performed using perturbation
   theory. When the effective theory is formulated as a lattice
   theory, the corresponding lattice Green's function is most easily
   evaluated non-perturbatively.  In the following we will refer to
   this procedure as matching the lattice and SM theories and $\mu_0$ as the matching scale.

   Before we go into the details of the lattice-SM matching, we start by 
   introducing the lattice methodology used to compute
   the local and the bilocal matrix elements. The evaluation of the $W$-$W$ diagrams will be described in detail as this is a new
   type of calculation.
   For the $Z$-exchange diagrams, we mainly focus on their difference from the 
   $\gamma$-exchange diagrams which dominate $K\to\pi\ell^+\ell^-$ decays and which have already been discussed in detail 
   in our previous paper~\cite{Christ:2015aha}.  

   \subsection{Evaluation of the matrix element of the local operator \boldmath{$Q_0$}}
   In this subsection we discuss the evaluation of $T_0\equiv\langle\pi^+\nu\bar{\nu}|Q_0^{\MS}(0)|K^+\rangle$, i.e. the matrix element of the 
   local operator $Q_0$.  The amplitude $T_0$ can be written as a product of 
   a hadronic matrix element and neutrino spinor wavefunctions:
   \begin{equation}
   T_0=Z_V\langle\pi^+|\bar{s}\gamma_\mu(1-\gamma_5)d(0)|K^+\rangle~\left[\bar{u}(p_\nu)\gamma_\mu(1-\gamma_5)v(p_{\bar{\nu}})\right].
   \end{equation}
   The charge-conserving hadronic factor can be related by an isospin rotation to the charge-changing matrix element 
   $\langle\pi^0|\bar{s}\gamma_\mu(1-\gamma_5)u|K^+\rangle$ which contains the hadronic effects in $K_{\ell 3}$ decay amplitudes. It can therefore be determined accurately 
   using precise measurements of $K^+\to\pi^0\ell^+\nu$ semileptonic decay amplitudes as input. In lattice QCD, the matrix element 
   $\langle\pi^+|\bar{s}\gamma_\mu(1-\gamma_5)d(0)|K^+\rangle$ can be determined by computing a three-point Euclidean correlation
   function. The matrix element of the axial-vector current vanishes because of parity symmetry and it is conventional to write the matrix element of the vector current in 
   terms of two invariant form factors:
   \begin{equation}
   \label{eq:SD_form_factor}
   Z_V\langle\pi^+|\bar{s}\gamma_\mu d(0)|K^+\rangle=i\cdot\left(f_+(-q^2)(p_K+p_\pi)_\mu+f_-(-q^2)(p_K-p_\pi)_\mu\,\right)\,,
   \end{equation}
   where $q=p_K-p_\pi$. 
   For negligible neutrino masses only the $f_+(-q^2)$ form factor contributes to $T_0$, so that
   \begin{eqnarray}
   \label{eq:T6}
   T_0=2i\cdot f_+(-q^2)~\left[\bar{u}(p_\nu){\slashed p}_K(1-\gamma_5)v(p_{\bar{\nu}})\right].
   \end{eqnarray}
   The $q^2$ dependence of the form factor $f_+(-q^2)$ can either be determined by a lattice QCD 
   calculation or provided by experimental measurement or indeed a combination of the two. For a recent lattice study and references to the original literature see Ref.\,\cite{Boyle:2015hfa}.  
   
   In lattice calculations, physical quantities are determined from the computation of multi-local correlation functions in Euclidean space. In this and the following sections of this paper, we use Euclidean conventions for the $\gamma$-matrices and momenta.   
   Thus for an on-shell particle with mass $m$, the Euclidean four-momentum $p=(p_0,\vec{p})$ is written as 
   $p=(iE,\vec{p})$ where $E=\sqrt{m^2+\vec{p}^{\,\,2}}$. Using this convention, $q^2>0$ ($q^2<0$) represents a 
   space-like (time-like) momentum transfer. The physical matrix elements are obtained from those defined using these Euclidean conventions by multiplying by the appropriate factors of $i$ as explained in detail in Appendix~\ref{sec:M-E}. This appendix also contains a full explanation of the notation we use for Euclidean quantities and the relations to the corresponding physical (Minkowski) ones.  The invariant form factors introduced in this paper, such as the $f_+$ and $f_-$ introduced in Eq.~\eqref{eq:SD_form_factor} will be defined consistently in both Euclidean and Minkowski space conventions.  This requires that minus signs be introduced when their arguments are expressed in terms of Euclidean four-vector dot products.

   \subsection{{\boldmath$W$-$W$} diagrams}
   \label{sec:WW}
   In this subsection we discuss elements of the calculation of the $W$-$W$ diagrams. We start in subsection\,\ref{subsubsec:scalar} by showing that the hadronic effects are contained in an invariant amplitude $F_{WW}$. In subsection\,\ref{subsubsec:exponential} we discuss the unphysical terms which increase exponentially in the length of the time integration range and how to subtract them. Such terms are generically present when evaluating the matrix elements of bilocal operators in Euclidean space whenever there are possible intermediate states of lower energy than the energy of the external states.

   \subsubsection{Extracting the scalar amplitude $F_{WW}$}
\label{subsubsec:scalar}
   \begin{figure}
   \centering
        \shortstack{\includegraphics[width=.3\textwidth]{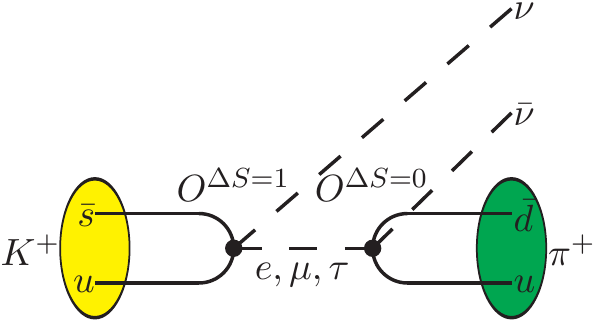}\\Type 1}
        \shortstack{\includegraphics[width=.3\textwidth]{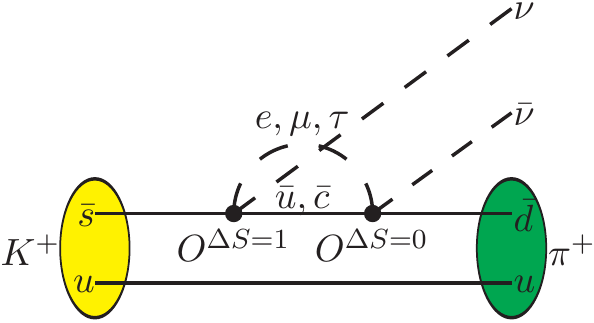}\\Type 2}
   \caption{Quark and lepton contractions for $W$-$W$ diagrams.}
   \label{fig:Wbox}
   \end{figure}

The hadronic effects in the contributions from $W$-$W$ diagrams to the decay amplitude are contained in the following matrix element of a bilocal operator:
   \begin{eqnarray}\label{eq:TWW}
   T_{WW}=\int d^4x\,\langle\pi^+\nu\bar{\nu}|T\{O_{u\ell}^{\Delta S=1}(x)\,
   O_{u\ell}^{\Delta S=0}(0)\}|K^+\rangle - \{u\to c\}\,.
   \end{eqnarray}
   The space-time location of $O_{u\ell}^{\Delta S=0}(y)$ defined in Eq.~(\ref{eq:B_WW}) 
   has been set at $y=0$ without loss of generality.
   The quark and lepton contractions for $T_{WW}$ are shown in Fig.~\ref{fig:Wbox}.
   The contraction between the two operators $O_{u\ell}^{\Delta S=1}$ and $O_{u\ell}^{\Delta S=0}$ produces an internal lepton propagator
   and the neutrino and anti-neutrino are emitted from the two different operators; the neutrino is emitted from $O^{\Delta S=1}$ at $x$ and the anti-neutrino from $O^{\Delta S=0}$ at the origin.  A Euclidean-space quantity such as that shown in Eq.~\eqref{eq:TWW} would normally be expressed directly as a Euclidean path integral.  Here we exploit the more compact Hilbert space notation for such a quantity.  It should be kept in mind that the time ordering represented by $T\{\ldots\}$ is required and that the time dependence of the operators is introduced by conjugation with the Euclidean time development operator $e^{-Ht}$ as described in Appendix~\ref{sec:M-E}.

In Appendix~\ref{sec:WW_scalar_amp} we show that $T_{WW}$ can be written in the form
   \begin{eqnarray}
   \label{eq:F_WW}
   T_{WW}=i\cdot F_{WW}(p_K,p_\nu,p_{\bar{\nu}})\,\left[\bar{u}(p_\nu){\slashed p}_K(1-\gamma_5)v(p_{\bar{\nu}})\right].
   \end{eqnarray}
   where $F_{WW}(p_K,p_\nu,p_{\bar{\nu}})$ is a scalar amplitude, which depends on 
   three of the independent external momenta $p_K$, $p_\pi$, $p_\nu$, $p_{\bar{\nu}}$.
   Since $F_{WW}(p_K,p_\nu,p_{\bar{\nu}})$ is Lorentz invariant, it can be written as a function of invariants:
   \begin{eqnarray}
   s=-(p_K-p_\pi)^2,\quad t=-(p_K-p_\nu)^2,\quad u=-(p_K-p_{\bar{\nu}})^2,
   \end{eqnarray}
   where $s+t+u=m_K^2+m_\pi^2$. 
   In a general $K_{l3}$ decay, it is convenient to study the differential decay 
   rate $d^2\Gamma/(ds\,d\cos\theta)$~\cite{bell1995quantum}, where $\theta$ is the angle between pion and 
   one of the neutrinos in the neutrino-pair rest frame. Following this convention,
   we choose the two independent variables as $s$ and $\Delta=u-t$.
   The former is the square of the invariant mass of the neutrino pair and the latter is proportional
   to $\cos\theta$. 

   \begin{figure}
   \centering
   \includegraphics[width=.8\textwidth]{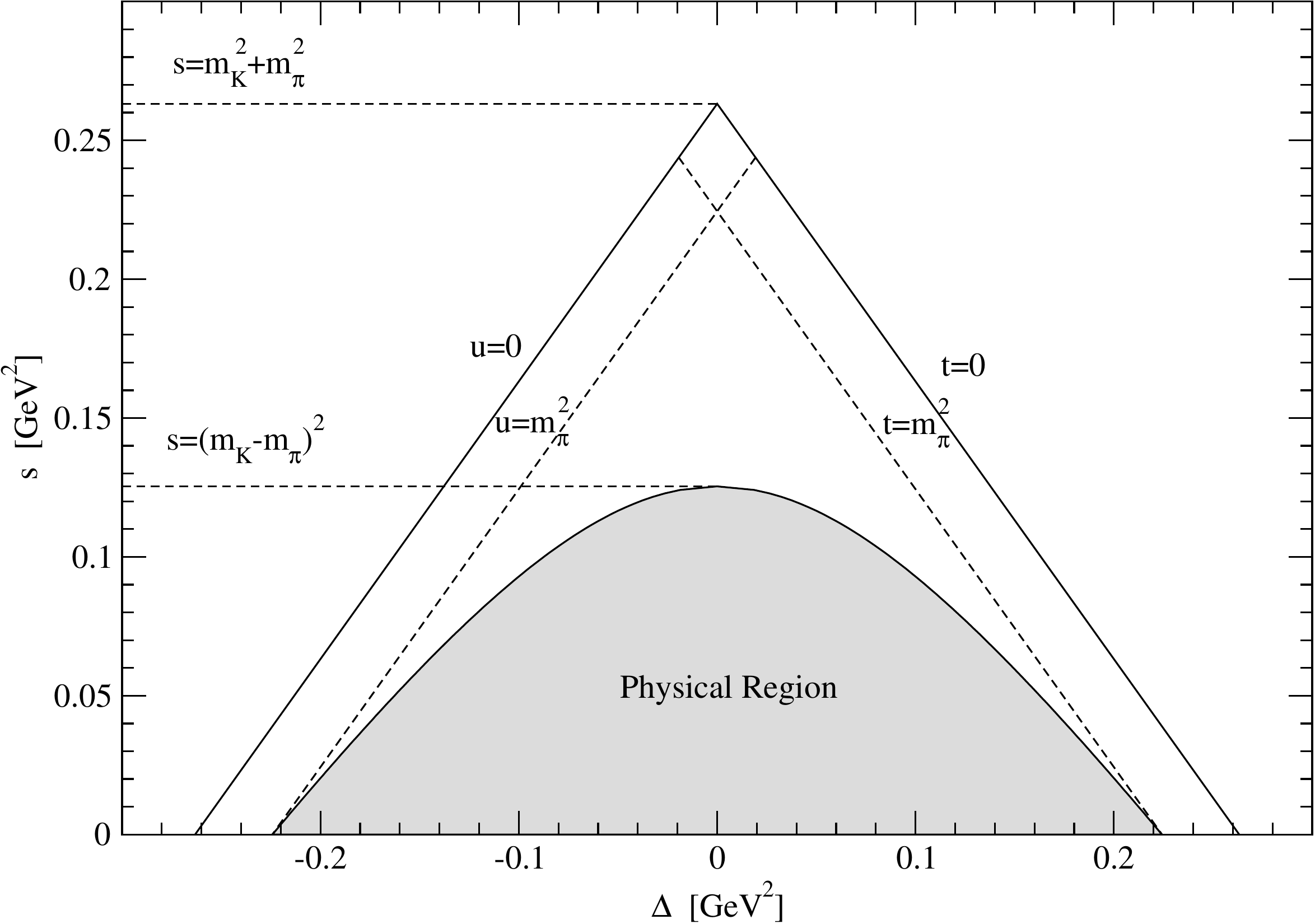}
   \caption{Dalitz plot for $K\to\pi\nu\bar{\nu}$.}
   \label{fig:Dalitz}
   \end{figure}

   To guarantee that the external particles are on shell,
   $s$ and $\Delta$ must be bounded by~\cite{Sehgal:1972wa}
   \begin{eqnarray}
   s\ge0 \quad \textrm{and}\quad\Delta^2\le (m_K^2+m_\pi^2-s)^2-4m_K^2m_\pi^2.
   \end{eqnarray}
   The physical range for $\{\Delta,s\}$ is shown in the Dalitz plot of Fig.~\ref{fig:Dalitz}.
   Note that in $K\to\pi\nu\bar{\nu}$ decays it is not practical to measure $\cos\theta$ experimentally.
   Therefore a differential decay rate $d\Gamma/ds$ is of more interest in phenomenology.
   Once the $\Delta$ dependence of $T_{WW}$ is determined, one can integrate  
   $\Delta$ over the physical phase space.

   \subsubsection{Unphysical terms growing exponentially with the Euclidean time integration range}
\label{subsubsec:exponential}
   \label{sec:exp_grow_contamination}
   In this subsection we study the  terms which grow exponentially as the time integration range is increased.  Such exponentially growing terms are a generic feature in the evaluation of integrals of matrix elements of bilocal operators over a large, but finite Euclidean time interval.  We note that this is the only unphysical consequence of evaluating such a bilocal operator in Euclidean space.  Here we consider specifically $\int d^4x\,\langle f|T[O^{\Delta S=1}(x)O^{\Delta S=0}(0)]|K\rangle$. We insert a complete set of states between the two operators and integrate over the Euclidean time region
   $-T_a<x_0<T_b$, where $T_a$ and $T_b$ are
   both positive.
   \begin{eqnarray}
    \int_{-T_a}^{T_b} dx_0\int d^3\vec{x}\,\langle f|T[O^{\Delta S=1}(x)O^{\Delta S=0}(0)]|K\rangle &
   \nn\\
   &\hspace{-1.2in}=\sum_{n_s}\cfrac{\langle f|O^{\Delta S=1}|n_s\rangle
   \langle n_s|O^{\Delta S=0}|K\rangle}{E_{n_s}-E_f}\left(1-e^{(E_f-E_{n_s})T_b}\right)
  \nn\\
  &\hspace{-0.8in} -\sum_n\cfrac{\langle f|O^{\Delta S=0}|n\rangle\langle n|O^{\Delta S=1}|K\rangle}{E_K-E_n}
   \left(1-e^{(E_K-E_n)T_a}\right).\hskip 0.2 in \label{eq:Euclidean}
   \end{eqnarray} 
   The two terms on the right hand side of Eq.~(\ref{eq:Euclidean}) come from the region $x_0>0$ and $x_0<0$ respectively.
   The states $|n\rangle$ and $|n_s\rangle$ represent non-strange and strangeness $S=1$ intermediate states respectively and include leptons as illustrated in Fig.\,\ref{fig:Wbox}.
   For the $K^+\to\pi^+\nu\bar{\nu}$ decay, the final state is given by $\langle f|=\langle\pi^+\nu\bar{\nu}|$.
   Since $|n_s\rangle$ are strange states, their energies
   $E_{n_s}$ are larger than $E_f=E_K$. Thus the exponential term $e^{(E_f-E_{n_s})T_b}$ vanishes
   at large $T_b$. However, the second term in Eq.~(\ref{eq:Euclidean}) still suffers from an 
   exponentially growing contamination at large $T_a$ if $E_n<E_K$.
   The lowest two intermediate states for $|n\rangle$ are given by a purely leptonic state $|\ell^+\nu\rangle$
   and a semi-leptonic state $|\pi^0\ell^+\nu\rangle$.
   As the energies of these intermediate states are lower than the energy of the initial state, the unphysical exponentially
   growing contamination must be removed from the Euclidean lattice calculation.
   In Appendix~\ref{sec:ground_state_WW} we give a detailed discussion on the removal of the
   exponentially growing contamination.
   The remaining contamination from other intermediate states, such as $|\pi\pi\ell^+\nu\rangle$ and
   $|3\pi\ell^+\nu\rangle$ are significantly suppressed by a phase-space factor as
   discussed in Sec.~\ref{sec:FV}. They can therefore be neglected.

   \subsection{{\boldmath$Z$-exchange} diagrams}

   \begin{figure}
   \centering
        \shortstack{\includegraphics[width=.3\textwidth]{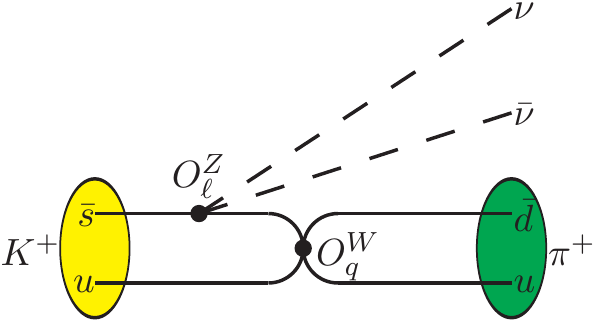}\\connected diag.}
        \shortstack{\includegraphics[width=.3\textwidth]{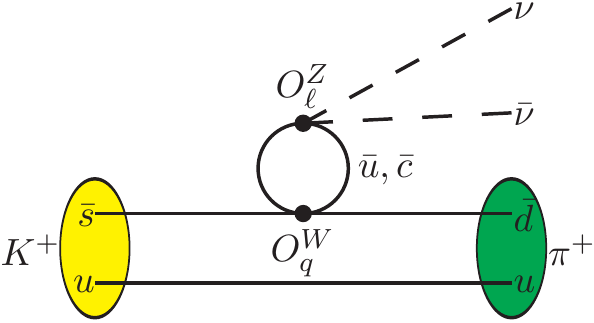}\\self-loop diag.}
        \shortstack{\includegraphics[width=.3\textwidth]{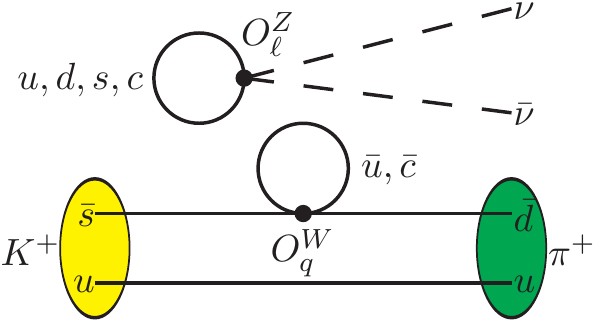}\\disconnected diag.}
   \caption{Samples of contractions contributing to $Z$-exchange diagrams. There are three different
   contraction structures: connected, self-loop and disconnected diagrams. For each case we show
   one example. A complete set of contractions can be found in our previous 
   publication~\cite{Christ:2015aha}.}
   \label{fig:Z_exchange}
   \end{figure}

In this subsection we discuss the evaluation of the 
$Z$-exchange diagrams. For these the neutrino and antineutrino are emitted from the same vertex
   and there is no internal lepton propagator.
   Examples of such diagrams for the 4-point correlation function are given in Fig.~\ref{fig:Z_exchange}. 
   We write the bilocal matrix element in the form
   \begin{eqnarray}
   T_Z&=&\int d^4x\,\langle\pi^+\nu\bar{\nu}|T[O_u^W(x)O_\ell^Z(0)]|K^+\rangle-\{u\to c\}
   \nn\\
   &=&T^Z_\mu\,\left[\bar{u}(p_\nu)\gamma_\mu(1-\gamma_5)v(p_{\bar{\nu}})\right],
   \end{eqnarray}
   where the hadronic part is defined as
   \begin{eqnarray}
   \label{eq:Z_hadronic}
   T^Z_\mu=\int d^4x\,\langle\pi^+|T[O_u^W(x) J_\mu^Z(0)]|K^+\rangle-\{u\to c\}.
   \end{eqnarray}
   The weak neutral current $J_\mu^Z$ has been defined in Eq.~\eqref{eq:JmuZ}.
   We separate $T^Z_\mu$ into two parts: $T^Z_\mu=T^{Z,V}_\mu+T^{Z,A}_\mu$, corresponding to the vector 
   ($V$) and axial-vector ($A$) components of $J_\mu^Z$.
   The $K\to\pi Z^*$ form factors are defined by 
   \begin{eqnarray}
   \label{eq:Z_form_factor}
   T_\mu^{Z,i}=i\cdot \left(F_+^{Z,i}(-q^2)(p_K+p_\pi)_\mu+F_-^{Z,i}(-q^2)q_\mu\right),\quad i=V,A,
   \end{eqnarray} 
   with $q=p_K-p_\pi$.
   Because the only possible Lorentz vectors are $p_K$ and $p_\pi$, the matrix element $T^{Z,i}_\mu$ must transform as a vector, not an axial-vector, under parity.
   This means that when calculating $T^{Z,i}_\mu$, we either keep the vector component of $J_\mu^Z$ with the parity-even component of $O_u^W$
   or the axial-vector component of $J_\mu^Z$ with the parity-odd component of $O_u^W$.
   The form factors $F_{\pm}^{Z,i}(-q^2)$ depend only on a single Lorentz invariant $q^2$.

   Since the spinor product $\bar{u}(p_\nu){\slashed q}(1-\gamma_5)v(p_{\bar{\nu}})$ vanishes for massless neutrinos,  $F_-^{Z,V}(-q^2)$ and $F_-^{Z,A}(-q^2)$ do not contribute to the amplitude.
   Only the form factors $F_+^{Z,i}(-q^2)$ are of interest.
   For the vector current, the Ward-Takahashi identity guarantees 
   $(m_K^2-m_\pi^2)F_+^{Z,V}(-q^2)=q^2F_-^{Z,V}(-q^2)$, so that there is only one independent form factor. For the axial-vector current, to separate $F_+^{Z,A}(q^2)$ from
   $T_\mu^{Z,A}$, we can compute the amplitude $T_\mu^{Z,A}$ for different Lorentz indices $\mu$. 
   This would require that either the kaon in the initial state
   or the pion in the final state should carry non-zero spatial momentum.

   As in the case of $T_{WW}$ a complete set of intermediate states can be inserted between
   $O_u^W$ and $J_\mu^Z$ in Eq.~(\ref{eq:Z_hadronic}).
   We need to remove the exponentially growing contamination for those intermediate states whose energies 
   are lower than that of the initial kaon. A detailed discussion of this subtraction for the case of the insertion of a vector current is 
    given in Ref.~\cite{Christ:2015aha}.  In that case the parity-odd intermediate states 
   $|\pi^+\rangle$ and $|3\pi\rangle$ will lead 
   to exponentially growing contamination which needs to be removed.
   For the axial-vector current insertion, the parity-even state $|2\pi\rangle$ will produce an exponentially
   growing contamination that also must be removed. Since we are only interested in $K^+$ decay, 
   the intermediate vacuum state does not
   contribute and the contribution of the $|2\pi\ell^+\nu\rangle$ ($K_{\ell 4}$) state is suppressed by phase space.

\section{Renormalization and short-distance correction}
\label{sec:SD}
In this section we discuss the renormalization of the ultraviolet divergences which appear in the calculation of the matrix elements of the bilocal operators introduced in Sec~\ref{sec:methodology}. This includes the standard renormalization of local composite operators which is discussed in the brief subsection~\ref{sec:operator renormalization}. Less standard is the presence of additional SD divergences which appear when the two local components of the bilocal operator approach each other. These additional ultra-violet divergences and their subtraction is discussed in detail in subsection~\ref{sec:SD_correction} which unsurprisingly makes up most of the section.

\subsection{Local operator renormalization}
\label{sec:operator renormalization}

   To produce the correct matrix element in the continuum limit, it is necessary (but not sufficient) for the lattice operators
   \{$O_{q\ell}^{\Delta S=1}$, $O_{q\ell}^{\Delta S=0}$\} for $W$-$W$ diagrams and
   \{$O^W_q$, $O^Z_\ell$\} for $Z$-exchange diagrams to
   be renormalized. We start by considering $O_{q\ell}^{\Delta S=1}$, $O_{q\ell}^{\Delta S=0}$ and
   $O^Z_\ell$ which are two-quark-two-lepton operators. 
   The leptonic current does not require renormalization and so we only need to deal with 
   the hadronic component which consists of vector and axial-vector currents. 
   In the massless quark limit, if the conserved vector and axial-vector currents  
   (in case of chiral lattice fermions, i.e. 
   domain wall or overlap fermions) are used, the Ward-Takahashi identity implies that the renormalization constants 
   $Z_V$ and $Z_A$ are equal to $1$. If instead, local currents are used then one needs to
   evaluate $Z_V$ and $Z_A$.
   The renormalization of the operators $Q_{1,q}$ and $Q_{2,q}$ (as well as $O_q^W$) has been discussed
   in our previous work~\cite{Christ:2015aha}.
   A more detailed description of the renormalization procedure can be found in 
   Refs.~\cite{Blum:2001xb,Blum:2011pu,Christ:2012se}.

   \subsection{Biocal operator renormalization}
   \label{sec:SD_correction}
   \begin{figure}
   \centering
   \includegraphics[width=.55\textwidth]{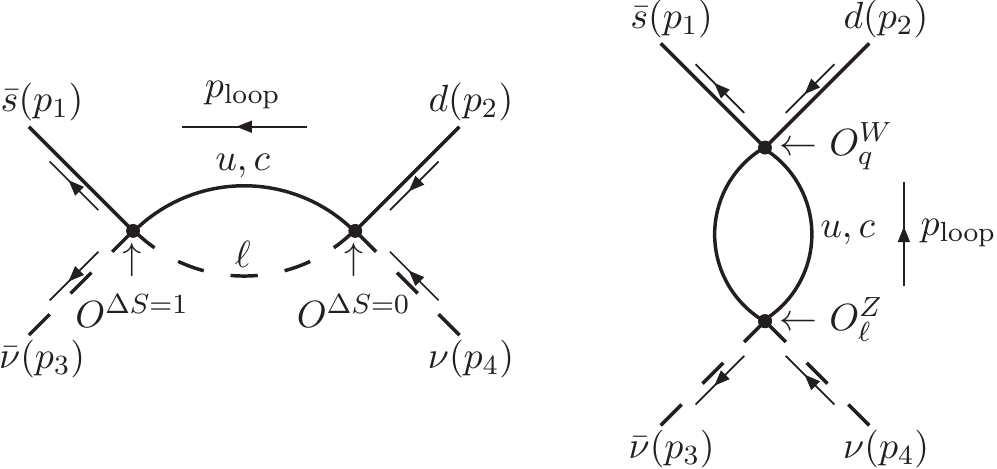}
   \caption{Left: SD divergent loop in $W$-$W$ diagrams. Right: SD divergent loop in $Z$-exchange diagrams.}
   \label{fig:SD_loop}
   \end{figure}

   In addition to the renormalization of the individual operators \{$O_{q\ell}^{\Delta S=1}$, 
   $O_{q\ell}^{\Delta S=0}$\} for $W$-$W$ diagrams and \{$O^W_q$, $O^Z_\ell$\} for $Z$-exchange diagrams,
   we need to consider possible new divergences which arise as the two operators approach each other, as shown in Fig.~\ref{fig:SD_loop}.
   Dimensional counting would allow for a potential quadratic divergence.
   In $W$-$W$ diagrams, the $V-A$ structure of the weak current and the GIM mechanism
   reduce the degree of divergence from quadratic to logarithmic since the leading divergence is independent of the quark mass.
   In $Z$-exchange diagrams, we imagine that $J_\mu^Z$ carries momentum $p=p_1-p_2=p_4-p_3$ (see Fig.\,\ref{fig:SD_loop})
   and recall that it
   contains both a vector and an axial-vector component.
   For the vector current insertion, if a conserved current is used, then
   the loop diagram is convergent and no lattice to continuum matching is required. This is 
   explained in Ref.~\cite{Isidori:2005tv} and in our previous paper~\cite{Christ:2015aha}.
   The situation is different for the insertion of the axial-vector current because the quark masses 
   $m_u$ and $m_c$ break the chiral symmetry explicitly. As a result, in addition to terms proportional to the tensors
   $p^2\delta_{\mu\nu}$ and $p_\mu p_\nu$, there are now terms proportional to $m_q^2\delta_{\mu\nu}$. 
   In all of these terms the degree of divergence is reduced by $2$, but now the remaining
   logarithmic divergence is not removed by the GIM mechanism since it contains terms proportional to $m_q^2$.
   Therefore, even if a conserved axial-vector current is
   used, the loop diagram shown in Fig.~\ref{fig:SD_loop} is still logarithmically divergent.
   This is the case for chiral lattice fermions for which the chiral symmetry is protected. 
   For Wilson fermions instead, where the chiral symmetry is violated by
   the Wilson term, then the GIM cancellation would lead to a linear divergence.
   We therefore propose to perform a lattice calculation of the $K^+\to\pi^+\nu\bar{\nu}$ decay amplitude 
   using domain wall fermions. As discussed above, whether a conserved or local axial-vector current 
   is used, we will need to deal with the logarithmic divergence remaining after the GIM cancellation from the SD region 
   where $O_\ell^Z$ and $O_q^W$ approach each other. 
   
   In the following subsections we present our proposed treatment of this additional SD divergence and the introduction of the counter term necessary to subtract it. We start however, with a description of the conventional approach, based on the perturbative evolution of the operators in the effective Hamiltonian to momentum scales below the mass of the charm quark and the non-perturbative evaluation of the matrix element of the remaining local operator(s). In this subsection we also explain why this is not the procedure which we propose to employ to determine the amplitudes for rare kaon decays.

\subsubsection{Perturbation theory calculations in the $\MS$ scheme}
\label{sec:PT}

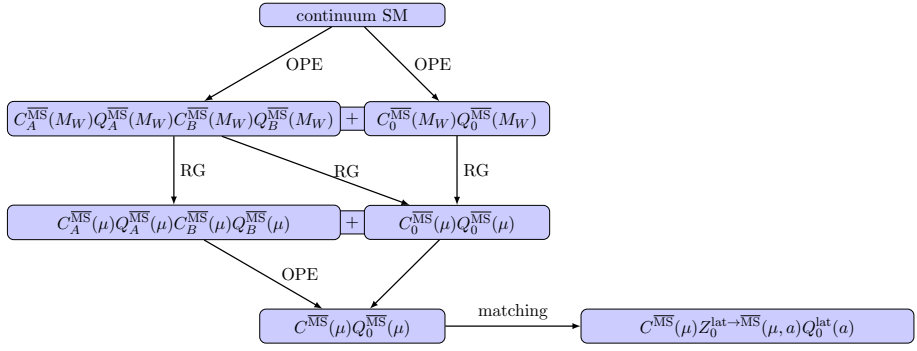
\begin{figure}
\begin{center}
\scalebox{0.6}{
\begin{tikzpicture}[node distance=2.3cm]
        \node[block]                  (init){continuum SM};
        \node[block, below of=init]   (nbrh){+};
        \node[block1, left=-1.8cm of nbrh]   (left){$C_A^{\MS}(M_W)Q^{\MS}_A(M_W)C_B^{\MS}(M_W)Q^{\MS}_B(M_W)$};
        \node[block, right=-1.8cm of nbrh]   (right){$C_0^{\MS}(M_W)Q^{\MS}_0(M_W)$};
        \node[block, below of=nbrh]   (ovgt){+};
        \node[block1, below of=left]  (left1){$C_A^{\MS}(\mu)Q^{\MS}_A(\mu)C_B^{\MS}(\mu)Q^{\MS}_B(\mu)$};
        \node[block, below of=right]  (right1){$C_0^{\MS}(\mu)Q_0^{\MS}(\mu)$};
        \node[block, below of=ovgt]   (ovgt1){$C^{\MS}(\mu)Q_0^{\MS}(\mu)$};
        \node[block1, right=3cm of ovgt1](rand){$C^{\MS}(\mu)Z^{\lat\to\MS}_0(\mu,a)Q_0^{\lat}(a)$};

        \path[line] (init) -- node[right]{~OPE}(right);
        \path[line] (init) -- node[right]{\ ~OPE}(left);
        \path[line] (left) -- node[right]{RG}(left1);
        \path[line] (left) -- node[right]{\ ~RG}(right1);
        \path[line] (right) -- node[right]{RG}(right1);
        \path[line] (left1) -- node[right]{\ ~OPE}(ovgt1);
        \path[line] (right1) -- (ovgt1);
        \path[line] (ovgt1) -- node[above]{matching}(rand);
      \end{tikzpicture}
}
\end{center}\caption{Schematic illustration of the steps in the treatment of the SD effects in perturbation theory. \label{fig:schematic}}\end{figure}

   We start by briefly reviewing perturbation theory calculations of the charm quark contribution to $K^+\to\pi^+\nu\bar{\nu}$ decays~\cite{Buchalla:1993wq,Buras:2005gr,Buras:2006gb}.  This is illustrated schematically by the diagram in Fig.\,\ref{fig:schematic}.  These considerations apply to each of the bilocal operators ${\mathcal B}_{WW}$ and ${\mathcal B}_Z$ given in Eqs.~\eqref{eq:B_WW} and ~\eqref{eq:B_Z}.  We will adopt a slightly generalized notation to allow us to discuss both cases at the same time.  Since the issues of operator renormalization and scale dependence are important, we also wish to explicitly show the Wilson coefficients, including their renormalization scale and scheme.  Thus, we will use the Wilson coefficient operator product $C_A Q_A$ to represent either the operator $O_{q\ell}^{\Delta S=1}$ ($W$-$W$ case) or $O_q^W$ ($Z$-exchange case).  As is shown in Eq.~\eqref{eq:O_W_operator}, for this second case we should actually write the sum of the product of two Wilson coefficients multiplying two operators.  In order to simplify our discussion we will ignore this familiar $2\times 2$ operator mixing complication (which is not difficult to treat) and use a single (coefficient)$\times$(operator) product in both cases.   Similarly we will use the product $C_B Q_B$ to represent either the operator $O_{q\ell}^{\Delta S=0}$ ($W$-$W$ case) or $O_\ell^Z$ ($Z$-exchange case).  Here $A$ and $B$ are generic labels for the four-fermion operators as indicated. The label $A$ should not be confused with the axial current.  In both cases the local counter term that must be introduced involves the same operator $Q_0$.  Thus we represent this local counter term by the product $C_0 Q_0$, where we should keep in mind that the Wilson coefficient $C_0$ will be different in the $W$-$W$ and $Z$-exchange cases.  We now describe each of the four steps in turn.

\textbf{Step 1:} The heavy $W$ and $Z$ bosons  are integrated out and the second-order weak interaction is written in a combination of a bilocal operator $\int d^4x~T[Q_A(x)Q_B(0)]^{\MS}(\mu)$ and a local operator $Q^{\MS}_0(\mu)$. Here $Q_{A,B}$ are local, four-fermion operators renormalized in the ${\MS}$ scheme. By setting up matching conditions at $\mu={\mathcal O}(M_W)$ and requiring the amplitude in the effective field theory to be the same as that in the full theory, one determines the coefficients $C_A^{\MS}(\mu)$, $C_B^{\MS}(\mu)$ and $C_0^{\MS}(\mu)$ at $\mu={\mathcal O}(M_W)$.  The local operator $Q_0^{\MS}(\mu)$ (and its Wilson coefficient $C_0^{\MS}(\mu)$) can be thought of as serving two closely-related purposes.   The first and most familiar is to represent phenomena, such as those that involve the top quark, which appear local below the scale of $M_W$.  The second purpose is to act as a counter-term removing the ultraviolet divergence from the SD region $x\approx 0$, where $Q_A(x)$ and $Q_B(0)$ approach each other.

\textbf{Step 2:}   As the next step the renormalization group equations are used to evolve the Wilson coefficients $C_A^{\MS}(\mu)$, $C_B^{\MS}(\mu)$ and $C_0^{\MS}(\mu)$ from the scale $\mu=M_W$ to lower scales. The evolution includes a mixing of the singular part of the bilocal operator $\int d^4x~T[Q_A(x)Q_B(0)]^{\MS}(\mu)$ into the local operator $Q_0^{\MS}(\mu)$. The corresponding renormalization group equations are an extension of those which govern the evolution of a set of local operators and are discussed in detail in 
Ref.\,\cite{Buchalla:1995vs}. The specific application to the rare kaon decays being studied here are described in Sec XI.B of \cite{Buchalla:1995vs}.

\textbf{Step 3:}   At the scale $\mu={\mathcal O}(m_c)$ we can perform a second Operator Product Expansion (OPE) and integrate out the active charm quark field. This can be done by evaluating the matrix element of the bilocal operator $T[Q_A(x) Q_B(0)]^{\MS}(\mu)$ and relating it to the matrix element of the local operator $\langle Q_0^{\MS}(\mu)\rangle$
   \begin{equation}
   \label{eq:bilocal_contribution}
   \int d^4x~\langle T[Q_A(x)Q_B(0)]^{\MS}(\mu) \rangle =r_{AB}^{\MS}(\mu)\langle Q_0^{\MS}(x=0,\mu)\rangle.
   \end{equation}
Following Refs.~\cite{Buchalla:1995vs,Buras:1998raa}, we use the term ``matrix element'' to mean ``amputated Green's functions of renormalized operators''.  Note that the corresponding LD contribution from the up quarks is suppressed by factors of $m_u^2/m_c^2$ (or $\Lambda_{\textrm{QCD}}^2/m_c^2$ from non-perturbative effects) relative to the terms that we are examining here at the energy scale $\mathcal{O}(m_c)$.  Of course, we must neglect such $\Lambda_{\textrm{QCD}}^2/m_c^2$ terms if Eq.~\eqref{eq:bilocal_contribution} is to reflect an underlying operator identity and the coefficient $r_{AB}$ to be independent of the ``amputated Green's functions of renormalized operators'' used to determine it.
   
 \textbf{Step 4:}  Finally, after integrating out the charm quark fields, the only remaining operator in the effective Hamiltonian is $C^{\MS}(\mu)Q_0^{\MS}(\mu)$, where the Wilson coefficient is given by
   \begin{equation}
   \label{eq:total_Wilson_coeff}
   C^{\MS}(\mu)=C_A^{\MS}(\mu)C_B^{\MS}(\mu)r_{AB}^{\MS}(\mu)+C_0^{\MS}(\mu).
   \end{equation}   
 At this stage the conventional approach is to calculate the $K^+\to\pi^+\nu\overline{\nu}$  matrix element of the local operator $Q_0^{\MS}(\mu)$. This can be done by starting with a lattice computation of the matrix element of $O^{\latt}(a)$ and then calculating the renormalization constant $Z_O^{\MS}(a\mu)$ to obtain the matrix element of $O^{\MS}(\mu)$. The renormalization constant $Z_O^{\MS}(a\mu)$ can either be calculated directly in perturbation theory or, as is now standard and generally more precise, to use \textit{non-perturbative renormalization} to obtain the operator in a scheme for which the renormalization conditions can be applied in a lattice calculation\,\cite{Martinelli:1994ty,Blum:2001xb,Sturm:2009kb} and then performing a continuum, perturbative matching calculation to obtain the operator in the $\MS$ scheme.
   
In this paper we propose an alternative approach in which steps 3 and 4 described above are not performed. The motivation for this is two fold.  First, we avoid using QCD perturbation theory at the charm quark scale where studies of the $K_L - K_S$ mass difference suggest poor convergence~\cite{Brod:2011ty}.  Second, we avoid relying on an effective theory in which the charm quark has been integrated out, which has further difficulties.  Once the charm quark has been integrated out, the higher-order corrections in the OPE are typically suppressed by powers of $\mu^2/m_c^2$. At this stage we are squeezed. On the one hand we would like to evolve to lower values of $\mu$ so that these omitted higher-order corrections are negligible and do not contribute large systematic uncertainties; on the other hand we cannot evolve the scale $\mu$ down to much lower values, e.g. $\mu=O(\Lambda_{\mathrm{QCD}})$, because perturbation theory surely fails at such low momentum scales. We propose instead, not to perform the second OPE (i.e. not to integrate out the charm quark) but to calculate directly the matrix elements of the bilocal operator $\int d^4x~T[Q_A(x)Q_B(0)]^{\MS}(\mu)$ and the local operator $Q_0^{\MS}(\mu)$ and combine them together to obtain the physical amplitude. 

This is the same approach that we have proposed to compute the LD contribution to the indirect CP violation parameter $\epsilon_K$~\cite{Christ:2012se, Bai:2015xxx}.  In contrast to the $K_L-K_S$ mass difference, for both $\epsilon_K$ and $K^+\to\pi^+\nu\overline{\nu}$ the second-order effective theory appropriate at the lattice scale of a few GeV contains logarithmic, ultra-violet divergences, requiring regulator-dependent counter terms.  In the case where a lattice regulator is to be used, extra steps are needed to determine these counter terms from those that are conventionally defined in the $\MS$ scheme.  In the following subsections, we will give a detailed description of our method in the current context.

   \subsubsection{The bilocal operator in the RI/SMOM scheme}
   To determine the matrix elements of bilocal and local operators renormalized in the $\MS$ scheme, we need first to adopt an intermediate scheme, which can be used in both non-perturbative 
   lattice QCD calculations as well as in continuum perturbation theory. Here we choose to use the RI/SMOM scheme. We consider the
   off-shell Green's functions with the four external legs carrying momenta: $\bar{s}(p_1)$, $d(p_2)$, $\bar{\nu}(p_3)$ and $\nu(p_4)$,
   as shown by Fig.~\ref{fig:SD_loop}. Since this Green's function is not a gauge-invariant observable, 
   the quark fields must be fixed in a particular gauge, e.g. the Landau gauge. 
   The ``non-exceptional'' external momenta $p_{1,2}$ are chosen to satisfy the condition 
   $p_1^2=p_2^2=(p_1-p_2)^2\gg\Lambda_{\mathrm{QCD}}^2$, which substantially suppresses the infra-red contamination in the computation 
   of the Green's function and hence improves the reliability of perturbation theory. A simple choice of $\{p_1,p_2\}$ is $p_1=(\xi,\xi,0,0)$ and $p_2=(\xi,0,\xi,0)$.  We define the
   RI/SMOM renormalization scale $\muRI$ by $\muRI^2\equiv p_{1,2}^2=2\xi^2$.  We emphasize that we have now introduced two distinct renormalization scales: the RI/SMOM renormalization scale $\muRI$ and the $\MS$ scale $\mu$.  While we could choose $\mu_0=\mu$, for generality and clarity of presentation we distinguish them here and below.

   Although the choice of neutrino momenta $p_3$ and $p_4$ is irrelevant 
   for the suppression of infra-red effects since no gluons connect to the neutrino lines, it does affect the momentum $p_{\mathrm{loop}}$
   flowing into the internal loop (see Fig.~\ref{fig:SD_loop}):
   \begin{eqnarray}
   p_{\mathrm{loop}}=\left\{
    \begin{array}{cl}
     p_1+p_3=p_2+p_4, & \textmd{for the $W$-$W$ diagram}, \\
     p_1-p_2=p_4-p_3, & \textmd{for the $Z$-exchange diagram}. \\
     \end{array}
     \right.   
   \end{eqnarray}
   For the $Z$-exchange diagram $p_{\mathrm{loop}}^2=\muRI^2$. For the $W$-$W$ diagram we can choose
   $p_3=(0,-\xi,0,-\xi)$ and $p_4=(0,0,-\xi,-\xi)$ which also leads to $p_{\mathrm{loop}}^2=\muRI^2$.
   Other choices of $\{p_3,p_4\}$ are also possible. For example
   if we interchange the definitions of $p_3$ and $p_4$, then $p_{\mathrm{loop}}^2=2\muRI^2$.
   What is required is that the neutrino momenta $p_3$ and $p_4$ are chosen such that
   $p_{\mathrm{loop}}$ is of the order of (or larger than) the renormalization scale $\muRI$ ($p^2_{\mathrm{loop}} \gtrsim \muRI^2$)
   so that the contributions to the momentum integrals $\int d^4p$ from regions of low momenta ($p^2\lesssim\Lambda_{\mathrm{QCD}}^2$) 
   are suppressed by one or more powers of $\Lambda_{\mathrm{QCD}}^2/p^2_{\mathrm{loop}}$. In this way, 
   we ensure SD dominance of the off-shell Green's function.

   Given the choice of external momenta $\{p_i\}$ described above, we can impose the RI/SMOM renormalization condition for the local
   operators $Q_A$, $Q_B$ and $Q_0$. Here we use the operator $Q_A$ to illustrate the procedure:
   \begin{equation}\label{eq:OlatttoRI}
   \langle Q_A^{\RI}(\mu_0) \rangle_{p_i^2=\mu_0^2}
   =[Z_q^{\RI}(\muRI)]^{-\frac{n}{2}}[Z_{O_A}^{\latt\to\RI}(a\muRI)]\langle Q_A^{\latt}(a) \rangle_{p_i^2=\mu_0^2}
   =\langle Q_A\rangle^{(0)}_{p_i^2=\mu_0^2}\,,
   \end{equation}
   where $\langle Q_A^{\RI} \rangle$ is the amputated Green's function of the renormalized operator $Q_A^{\RI}(\muRI)$,
   $\langle Q_A^{\latt} \rangle$ is the amputated Green's function of the bare lattice operator $Q_A^{\latt}(a)$ and $\langle Q_A \rangle^{(0)}$ is the 
   tree-level amputated Green's function. The subscripts $p_i^2=\mu_0^2$ in Eq.\,(\ref{eq:OlatttoRI}) indicate that the Green's functions are evaluated with the choice of momenta described above, i.e. with $p_1^2=p_2^2=(p_1-p_2)^2=\mu_0^2$.
   $Z_q$ is the quark's wave function renormalization constant; see Ref.~\cite{Sturm:2009kb} for the detailed definitions to be used in the RI-SMOM schemes and $n$ is the number of external quark lines. For the rare kaon decays being studied here $n=2$ and below we shall simply replace $n$ by 2.
   The renormalization constant $Z_{Q_A}^{\lat\to\RI}(a\muRI)$ relates the renormalized operator 
   $Q_A^{\RI}(\muRI)$ and the bare operator $Q_A^{\latt}(a)$ through
   the relation $Q_A^{\RI}(\muRI)=Z_{Q_A}^{\lat\to\RI}(a\muRI)Q_A^{\latt}(a)$. It can be determined non-perturbatively by 
   evaluating $\langle Q_A^{\latt} \rangle$ with the given external momentum $\{p_i\}$ and imposing the condition in Eq.\,(\ref{eq:OlatttoRI}). 
   
   As the next step, one can calculate the conversion factor
   $Z_{Q_A}^{\RI\to\MS}(\mu/\mu_0)$ perturbatively, relating the renormalized operators in the RI/SMOM and $\MS$ schemes through
   $Q_A^{\MS}(\mu)=Z_{Q_A}^{\RI\to\MS}(\mu/\mu_0)Q_A^{\RI}(\mu_0)$. Using the conversion factor $Z_{Q_A}^{\RI\to\MS}(\mu/\mu_0)$ and the renormalization constant
   $Z_{Q_A}^{\latt\to\RI}(a\muRI)$, the $\MS$ operator can be related to the bare lattice operator through
   $Q_A^{\MS}(\mu)=Z_{Q_A}^{\RI\to\MS}(\mu/\mu_0)Z_{Q_A}^{\latt\to\RI}(a\muRI)Q_A^{\lat}(a)\equiv Z_{Q_A}^{\MS}(a\mu)Q_A^{\latt}(a)$.

Next we extend the RI/SMOM scheme to provide a regularization-independent definition of the bilocal product of $Q_A$ and $Q_B$.  Here will we use the notation:
\begin{equation}
\{Q_A^S Q_B^S\}^{S'}(y) = \int d^4 x\, T\bigl\{Q_A^S(x) Q_B^S(y)\bigr\}^{S'},
\label{eq:bilocal_convention}
\end{equation}
where $S$ indicates the scheme used to define the local operators $O_A$ and $O_B$ while  $S'$ labels the method used to define the singularity when $x =y$.  Here the labels $S$ and $S'$ can be a combination of the three choices $\MS$,  $\lat$ or $\RI$.  For simplicity we will usually choose $y=0$ and not show this argument explicitly.  While the choices $S'=\MS$ and $\lat$ are defined by standard conventions, the case $S'=\RI$ is defined by imposing the condition:
  \begin{equation}
   \label{eq:bilocal_renormalizaton_cond_gen}
   \bigl\langle \{Q_A^S Q_B^S\}^{\RI}_{\muRI}\bigr\rangle_{p_i^2=\mu_0^2} = 0\,,
   \end{equation}
where the subscript ${p_i^2=\mu_0^2}$ indicates the amputated, four-Fermi Green's function evaluated for the non-exceptional external momenta described above.  The subscript $\mu_0$ added to the bilocal operator itself indicates the scale dependence that this $\RI$ operator has acquired because of the condition used to define it.

To relate the bilocal operators $\{Q_A^{\RI} Q_B^{\RI}\}^{\RI}_{\muRI}$ and $\{Q_A^{\RI} Q_B^{\RI}\}^{\lat}_a$, we can write
   \begin{equation}\label{eq:O1O2RI}
   \{Q_A^{\RI} Q_B^{\RI}\}^{\RI}_{\muRI} = \{Q_A^{\RI}(\muRI)Q_B^{\RI}(\muRI)\}^{\lat}_a 
                -X_{AB}(\muRI,a) \,Q_0^{\RI}(\muRI)\,,
   \end{equation}
where the last term on the right-hand side is introduced to compensate for the different treatment of the singularity in the product $Q_A(x) Q_B(0)$ as $x\to 0$ in the two different schemes.  Although each of the renormalized local operators $Q_A^{\RI}$, $Q_B^{\RI}$ and $Q_0^{\RI}$ individually are independent of the ultraviolet cut-off $a$, the additional SD divergence in $\{Q_A^{\RI}Q_B^{\RI}\}^{\lat}$ is regulated using the lattice cut-off. The coefficient $X_{AB}(\mu_0,a)$ therefore has a dependence on $a$ and is defined by the subtraction condition in Eq.~\eqref{eq:bilocal_renormalizaton_cond_gen}:
   \begin{equation}
   \label{eq:bilocal_renormalizaton_cond_lat}
   \bigl\langle \{Q_A^{\RI} Q_B^{\RI}\}^{\RI}_{\muRI}\bigr\rangle_{p_i^2=\mu_0^2}=\bigl\langle \{Q_A^{\RI}Q_B^{\RI}\}_a^{\lat}\bigr\rangle_{p_i^2=\mu_0^2} - X_{AB}(\muRI,a) \bigl\langle Q_0^{\RI}(\mu_0)\rangle_{p_i^2=\mu_0^2}=0\,.
   \end{equation}
These Green's functions are calculated by computing the corresponding Green's functions for the bare lattice operators and multiplying by the $Z^{\latt\to\RI}$ renormalization constant for each of the local operators. Using the renormalization condition~(\ref{eq:bilocal_renormalizaton_cond_lat}) we can determine the coefficient $X_{AB}(\muRI,a)$ non-perturbatively and hence can define the RI/SMOM bilocal operator $\{Q_A Q_B\}^{\RI}_{\muRI}$ through Eq.~\eqref{eq:bilocal_renormalizaton_cond_lat} with no ambiguity and no dependence on $a$. 

   Finally we can express the $\MS$ bilocal operator in terms of the RI/SMOM bilocal and an additional local operator by using the analogous equation to Eq.~\eqref{eq:O1O2RI},
   \begin{equation}\label{eq:O1O2MS}
   \{Q_A^{\MS} Q_B^{\MS}\}^{\MS}_{\mu}=Z_{Q_A}^{\RI\to\MS}(\mu/\muRI) Z_{Q_B}^{\RI\to\MS}(\mu/\muRI) \{Q_A^{\RI} Q_B^{\RI}\}^{\RI}_{\muRI}+Y_{AB}(\mu,\muRI)\,Q_0^{\RI}(\muRI).
   \end{equation}
Green's functions of the bilocal operator $\{Q_A^{\MS} Q_B^{\MS}\}^{\MS}_\mu$ are evaluated using dimensional regularization of all the ultraviolet divergences and their subtraction following the standard procedure to define the $\MS$ scheme. The $\mu$-dependence of such Green's functions has contributions not only from the anomalous dimensions of $Q_A $ and $Q_B$ (and reproduced by the first term on the left-hand side of Eq.\,(\ref{eq:O1O2MS})) but also from the SD region and contained in the coefficient $Y_{AB}(\mu,\muRI)$. To determine $Y_{AB}(\mu,\muRI)$ we calculate the amputated Green's functions for both sides of Eq.\,(\ref{eq:O1O2MS}) at $p_i^2=\muRI^2$ and impose the renormalization condition Eq.\,(\ref{eq:bilocal_renormalizaton_cond_gen}) so that:
   \begin{eqnarray}
  \bigl\langle \{Q_A^{\MS} Q_B^{\MS}\}^{\MS}_\mu\bigr\rangle_{p_i^2=\muRI^2} &&
   \nn\\
 && \hskip -1.1 in  =\frac{Z_q^{\RI}(\muRI)}{Z_q^{\MS}(\mu)}\left[Z_{Q_A}^{\RI\to\MS}(\mu/\muRI)Z_{Q_B}^{\RI\to\MS}(\mu/\muRI)\bigl\langle \{Q_A^{\RI} Q_B^{\RI} \}^{\RI}_{\muRI^2}\bigr\rangle_{p_i^2=\muRI^2}+Y_{AB}(\mu,\muRI) \bigl\langle Q_0^{\RI}\bigr\rangle_{p_i^2=\muRI^2}\right]
\nn\\
 && \hskip -1.1 in  =  \frac{Z_q^{\RI}(\muRI)}{Z_q^{\MS}(\mu)}\,Y_{AB}(\mu,\muRI)\langle Q_0\rangle^{(0)}_{p_i^2=\muRI^2}\,,
   \end{eqnarray}
where the superscript $(0)$ denotes \textit{tree-level}, and reminds us that the RI/SMOM renormalisation condition is $\langle Q_0^{\textrm{RI}}\rangle_{p_i^2=\mu_0^2}=
\langle Q_0\rangle^{(0)}_{p_i^2=\mu_0^2}$.
 In this way we can determine the coefficient $Y_{AB}(\mu,\muRI)$ and hence, using Eq.~\eqref {eq:O1O2MS}, express the bilocal operator $\{Q_A Q_B\}^{\MS}(\mu)$ in terms of operators that are defined in lattice QCD.

\subsubsection{Numerical strategy for bilocal operator renormalization}

As reviewed in Sect.~\ref{sec:PT}, electroweak and QCD perturbation theory can be used to determine a combination of bilocal and local operators, defined in the $\MS$ scheme at a scale $\mu$, whose matrix element between $K^+$ and $\pi^+\nu\overline{\nu}$ states will accurately determine the rare $K^+\to\pi^+\nu\overline{\nu}$ decay amplitude, provided the scale $\mu$ is sufficiently large that QCD perturbation is accurate.  Following Eq.~\eqref{eq:ct}  we can write this second order weak operator, before the final integral over space time, as the combination:
\begin{eqnarray}
\mathcal{B}^{\MS}_{WW}(y) +  \mathcal{B}^{\MS}_Z(y) + C_0^{\MS} Q_0^{\MS}(y).
\label{eq:2nd_order_decay_op}
\end{eqnarray}
When the $\MS$ scale $\mu$ is below the bottom quark mass, one expects that the largest contribution come from the second, $C_0^{\MS} Q_0^{\MS}$ term in this operator since it contains a $\ln(M_W/m_b)$ factor which the bilocal operators $\mathcal{B}^{\MS}_{WW}(y)$ and $\mathcal{B}^{\MS}_Z(y)$ do not.  The contribution of this local term to the $K^+\to\pi^+\nu\overline{\nu}$ decay rate can be accurately computed and the achieved accuracy of this computation underlies the experimental and theoretical interest in this process.   

In this paper we wish to augment this capability with a first-principles calculation of the matrix elements of the bilocal operators in Eq.~\eqref{eq:2nd_order_decay_op}.  To the extent that this term is relatively small, our methods do not need to be as precise as those used to determine the matrix element of the local operator.  For example, we may be able to obtain a useful result if we employ only leading-order formulae for the perturbative coefficients $Y(\mu,\muRI)$ which relate the $\MS$-normalized bilocal operators appearing in Eq.~\eqref{eq:2nd_order_decay_op} and the RI-normalized bilocal operators which can be evaluated non-perturbatively using lattice methods.  As we increase the scale $\mu$ appearing in Eq.~\eqref{eq:2nd_order_decay_op}, the use of QCD perturbation theory to determine the Wilson coefficients appearing in that equation will become more reliable.  However, this will also cause the contribution of the bilocal operator to increase, requiring a higher precision from the lattice calculation if the over-all error is to decrease.

We will make the preceding discussion concrete by writing out an explicit example expressing the perturbatively-determined operator $\mathcal{B}^{\MS}_Z(y)$ in terms of operators and coefficients that can be determined directly from a lattice QCD calculation:
\begin{eqnarray}
 \mathcal{B}^{\MS}_{Z,A} &=& 
\Biggl\{\Bigl(C_1(\mu)^{\MS} Q_{1,u}^{\MS} + C_2(\mu)^{\MS} Q_{2,u}^{\MS}\Bigr)
\Bigl( J^{A}_\mu\overline{\nu}\gamma^\mu(1-\gamma^5)\nu\Bigr)-\{u\to c\} \Biggr\}_\mu^{\MS} \nn\\
&=& 
\Biggl\{\Bigl(\sum_{i,j=1,2} C_i(\mu)^{\MS} Z_{ij}^{\RI\to\MS} Q_{j,u}^{\RI}\Bigr)
\Bigl(J^A_\mu\overline{\nu}\gamma^\mu(1-\gamma^5)\nu\Bigr)-\{u\to c\} \Biggr\}_{\muRI}^{\RI} \nn \\
&&\hskip 0.5 in + \sum_{i=1,2} C_i^{\MS}(\mu) Y_{Q_i,J^A}(\mu,\muRI)Q_0^{\RI}(\muRI), 
\label{eq:example}
\end{eqnarray}
where we have considered the case of the operator $O_q^W$ defined in Eq.~\eqref{eq:O_W_operator} and included the required operator mixing but examined only the hadronic axial current component of the current $J_\mu^Z$ given in Eq.~\eqref{eq:JmuZ}. 

\section{Finite-volume effects}
\label{sec:FV}
   When second-order weak amplitudes that involve multi-particle intermediate states are computed in finite volume, potentially significant finite-volume corrections can appear.
References~\cite{Christ:2010gi,Christ:2014qaa,Christ:2015pwa} give detailed formulae which determine the finite-volume (FV) correction for the case of the two-pion intermediate state that appears in a calculation of the $K_L$-$K_S$ mass difference. The same approach can be used to determine FV effects in rare kaon decay amplitudes.
   The finite volume effects discussed in this section and in the
   above references are those which fall as powers of the lattice size
   and arise from the degeneracy between possible intermediate states
   and the initial and final states in the process being considered.  Here
   we do not address the presumably smaller FV effects
   which fall exponentially as the volume increases. 

As is well-known, power-law, FV corrections are related to the on-shell amplitudes
   $A(K\to\{n\})$, where $\{n\}$ represents an intermediate state made up of $n$ particles.
   As more particles are included in $\{n\}$, we expect that the FV correction
   will be increasingly suppressed by the resulting reduced phase-space.
   In Table~\ref{tab:phase_space_factor} we list the relevant braching ratios 
   of $K\to\{n\}$ from the  Particle Data Group~\cite{Agashe:2014kda}. Since the
   $K_{e2}$ decay is helicity suppressed, we can compare the other entries
 in Table~\ref{tab:phase_space_factor} with that for $K_{\mu2}$ 
   to estimate the effect of this phase-space suppression.
   As the number of daughter particles increases, the braching ratios are significantly suppressed.
   The only exception is seen in the comparison between the decay modes $K^+\to\pi^+\pi^0$ and $K^+\to3\pi$, 
   where the branching ratio is only 3 times smaller in $K^+\to3\pi$ decay. 
   However, this is because only the $I=2$ pion-pion state contributes to the $K^+\to\pi^+\pi^0$ mode and the
   corresponding decay amplitude is highly suppressed because of 
   the $\Delta I=1/2$ rule as explained in Ref.~\cite{Boyle:2012ys}.
   If we consider instead the neutral kaon decays, to which the $I=0$ pion-pion state also contributes, and compare the
   decay width between $K_S\to2\pi$ and $K_L\to3\pi$, a large 
   phase-space suppression can be observed in Table~\ref{tab:phase_space_factor}.

\begin{table}
  \centering
  \begin{tabular}{l|c|c}
    \hline
    $K\to\{n\}$ & Branching ratio & relevant diagrams\\
    \hline
    $K^+\to\mu^+\nu_\mu$ &  $6.355(11)\times10^{-1}$ & \multirow{2}{*}{$W$-$W$ diagram}\\
    $K^+\to2\pi\mu^+\nu_\mu$ & $4.254(32)\times10^{-5}$  & \\
    \hline
    $K^+\to\pi^0e^+\nu_e$ & $3.353(34)\times10^{-2}$  & \multirow{2}{*}{$W$-$W$ diagram}\\
    $K^+\to3\pi e^+\nu_e$ & $<3.5\times10^{-6}$ & \\
    \hline
    $K^+\to\pi^+\pi^0$ & $2.066(8)\times10^{-1}$ & $Z$-exchange diagram, $J_\mu^{Z,A}$ \\
    $K^+\to3\pi$ & $7.35(5)\times10^{-2}$ & $Z$-exchange diagram, $J_\mu^{Z,V}$ \\
    \hline\hline
    $K\to\{n\}$ & Decay width [eV] & relevant diagrams\\
    \hline
    $K_S\to2\pi$ & $7.343(13)\times10^{-6}$   & $Z$-exchange diagram, $J_\mu^{Z,A}$ \\
    $K_L\to3\pi$ & $4.125(30)\times10^{-9}$    & $Z$-exchange diagram, $J_\mu^{Z,V}$ \\
    \hline
  \end{tabular}
  \caption{Branching ratios and decay widths for $K\to\{n\}$ decays. The third column
   gives the relevant diagrams to which the $K\to\{n\}$ amplitudes contribute.
   As $n$ increases, a large suppression can be observed in the $K^+\to\{n\}$ branching ratio. 
   The only exceptions to this trend ($K^+\to\pi^+\pi^0$ and
   $K^+\to3\pi$ decays) can be explained by the $\Delta I=1/2$ rule. In the neutral kaon decay, we
   show the suppression of the decay width from $K_S\to2\pi$ to $K_L\to3\pi$ decay. Here the decay width
   is given in units of eV.}
  \label{tab:phase_space_factor}
\end{table}

   From Table~\ref{tab:phase_space_factor}, we conclude that for the $W$-$W$ diagrams, we may neglect the FV effects associated 
   with on-shell $K^+\to2\pi\ell^+\nu_\ell$ and $K^+\to3\pi\ell^+\nu_\ell$ amplitudes,
   which are highly phase-space suppressed.
   We need to
   consider only the FV corrections related to $K^+\to\ell^+\nu_\ell$ and
   $K^+\to\pi^0\ell^+\nu_\ell$ amplitudes. Here, the 4-momentum of the intermediate neutrino is completely
   determined by the $\langle\pi^+\nu\bar{\nu}|$ final state. Therefore, no power-law,
   FV effects exist for the $|\ell^+\nu_\ell\rangle$ intermediate state.
   For the state $|\pi^0\ell^+\nu_\ell\rangle$, 
   the corresponding FV correction, $T_{WW}^{FV}=T_{WW}(L)-T_{WW}(\infty)$,
   can be expressed as
   \begin{eqnarray}
   \label{eq:WW_FV}
   T_{WW}^{FV}&=&\left(\frac{1}{L^3}\sum_{\vec{k}}\int \frac{dk_0}{2\pi}
   -{\mathcal P}\int\frac{d^4k}{(2\pi)^4}\right)
   \nn\\
   &&\hspace{1cm}
   \left\{A_\alpha^{K^+\to\pi^0}(p_K,k)\frac{1}{k^2+m_\pi^2}
    A_\beta^{\pi^0\to\pi^+}(k,p_\pi)\right\}
   \nn\\
   &&\hspace{1.5cm}\times
   \left\{\bar{u}(p_\nu)\gamma^\alpha(1-\gamma_5)
   \frac{i({\slashed P}-{\slashed k})+m_{\bar{\ell}}}{(P-k)^2+m_{\bar{\ell}}^2}
   \gamma^\beta(1-\gamma_5)v(p_{\bar{\nu}})\right\},
   \end{eqnarray}
   where $k$ is the momentum carried by the intermediate $\pi^0$ and $P=p_K-p_\nu$ is the total
   momentum flowing into the $\pi^0$-$\ell^+$ loop. 
   The second line of Eq.~(\ref{eq:WW_FV}) corresponds to the
   sequence of hadronic transitions $K^+\to\pi^0\to\pi^+$. The $K^+\to\pi^0$
   and $\pi^0\to\pi^+$ transition amplitudes are given by
   \begin{eqnarray}
   \label{eq:trans_amp}
   A_\alpha^{K^+\to\pi^0}(p_K,k)&=&Z_V\langle\pi^0(k)|\bar{s}\gamma_\alpha u(0)|K^+(p_K)\rangle,
   \nn\\
   A_\beta^{\pi^0\to\pi^+}(k,p_\pi)&=&Z_V\langle\pi^+(p_\pi)|\bar{u}\gamma_\beta d(0)|\pi^0(k)\rangle.
   \end{eqnarray}
   Though the intermediate $\pi^0$ can carry an off-shell momentum, only the on-shell
   $K^+\to\pi^0$ and $\pi^0\to\pi^+$ amplitudes can contribute to $T_{WW}^{FV}$. 
   Therefore in Eq.~(\ref{eq:trans_amp}) we simply
   define $A_\alpha^{K^+\to\pi^0}(p_K,k)$ and $A_\beta^{\pi^0\to\pi^+}(k,p_\pi)$ using the on-shell 
   pion state $|\pi^0\rangle$. To estimate the FV correction, we need to evaluate these transition 
   amplitudes in our lattice calculation. Once available, these amplitudes can also be
   used to remove the exponentially growing contamination since the $|\pi^0\ell^+\nu\rangle$ state
   possibly has a lower energy than the initial kaon.
   The third line of Eq.~(\ref{eq:WW_FV}) gives the leptonic contribution which
   involves a lepton propagator.

   Although the expression in Eq.~(\ref{eq:WW_FV}) is complicated, we can write it in
   a simpler but more general form as 
   \begin{eqnarray}
   I_{FV}=I(L)-I(\infty)=\left(\frac{1}{L^3}\sum_{\vec{k}}\int \frac{dk_0}{2\pi}
   -{\mathcal P}\int\frac{d^4k}{(2\pi)^4}\right)\frac{f(k_0,\vec{k})}{(k^2+m_1^2)((P-k)^2+m_2^2)}.
   \end{eqnarray}
For the case $\vec P =0$, this expression can be evaluated using formulae given in Ref.~\cite{Christ:2015pwa}, simplified by the vanishing of the $\pi^0$-$\ell^+$ scattering phase shift, since we are not including electromagnetic effects.  However, for $\vec P\ne 0$ this discussion must be generalized following the treatment given by Kim, Sachrajda and Sharpe in Ref.~\cite{Kim:2005gf} for the case $m_1=m_2$, boosting the system into the center-of-mass frame. For $m_1\neq m_2$, a similar result is given in Ref~\cite{Hansen:2012tf}.  We conclude that if the hadronic transition amplitudes $A_\alpha^{K^+\to\pi^0}(p_K,k)$ and $A_\beta^{\pi^0\to\pi^+}(k,p_\pi)$ have been determined, one can evaluate the FV correction $T_{WW}^{FV}$ using known methods.

   For the $Z$-exchange diagrams, the FV effect resulting from the transition $K^+\to3\pi$ is significantly suppressed by a
   phase-space factor, and that related to $K^+\to\pi^+\pi^0$ is suppressed by 
   $\Delta I=1/2$ rule.
   Therefore, we can choose to neglect both of these sources of finite volume error in a near-term lattice calculation. 
   If we wish to have a more accurate understanding of how small these FV corrections may be,
   we can evaluate the larger FV piece coming from the $\pi^+\pi^0$ intermediate state.  Since the momenta for three non-interacting particles
   in the $\langle\pi^+\nu\bar{\nu}|$  final state are assigned explicitly, no power-law, FV effect of the sort identified by Lellouch and L\"uscher~\cite{Lellouch:2000pv} is present for this rare kaon decay.  We can then treat $\langle\pi^+\nu\bar{\nu}|$ as a single-particle
   state $\langle \widetilde{\pi}^+|$ and again extend the FV correction formula derived for the case 
   of the $K_L-K_S$ mass difference~\cite{Christ:2015pwa} to the rare kaon decay. In this way, we obtain the FV correction
   \begin{eqnarray}
   \label{eq:FV_KLKS}
   &&\sum_n\frac{\langle \widetilde{\pi}^+|{O}^{Z}_\ell|n\rangle^{FV}
   {}^{FV}\langle n|O^W_q|K^+\rangle}{m_K-E_n}
   -{\mathcal P}\int_{2m_\pi}^\infty dE\sum_\alpha
   \frac{\langle\widetilde{\pi}^+|O^Z_\ell|\alpha,E\rangle^\infty
   {}^\infty\langle\alpha,E|O^W_q|K^+\rangle}{m_K-E}
   \nn\\
   &=&\cot(\phi(E)+\delta(E))\frac{d(\phi(E)+\delta(E))}{dE}
   \bigg|_{E=m_K}\langle \widetilde{\pi}^+|O^Z_\ell|\pi^+\pi^0,m_K\rangle^{FV}
   {}^{FV}\langle\pi^+\pi^0,m_K|O^W_q|K^+\rangle.
   \nn\\
   \end{eqnarray} 
   Here we use the notation of Ref.~\cite{Christ:2015pwa}. Making the replacement 
   $\langle \widetilde{\pi}^+|O^{Z}_\ell\,  \to \, \langle\pi^+|J_\mu^Z$
   in Eq.~(\ref{eq:FV_KLKS}), we obtain the FV correction formula for $T_\mu^Z$.

\section{Conclusion}
\label{sec:conclusion}
   With the development of new methods~\cite{Christ:2010gi,Christ:2012np, Christ:2012se, Christ:2014qaa,Bai:2014cva, Christ:2015pwa},   it is now possible to calculate the long-distance
   contributions to second-order weak amplitudes, such as the $K_L-K_S$ mass difference $\Delta M_K$ and $\epsilon_K$, directly using lattice QCD.  These methods have now been extended in Ref.~\cite{Christ:2015aha} to address the long-distance contributions to the rare kaon decay $K\to\pi\ell^+\ell^-$.  The present paper is a companion to Ref.~\cite{Christ:2015aha}, focusing here on developing lattice methods that can be used to compute the long-distance corrections to the rare kaon decay $K\to\pi\nu\bar{\nu}$.  In each  of these treatments, those contributions which are identified as long-distance and targeted by the proposed lattice methods include all energy scales at or below an energy that is conservatively chosen to exceed the charm quark mass.  Thus, these methods will allow calculations in which QCD perturbation theory is used only at energy scales which lie above the charm quark mass. 

   Since the NA62 experiment at CERN is now collecting data for $K^+\to\pi^+\nu\bar{\nu}$ 
   and the KOTO experiment at J-PARC in Japan is designed to search for the $K_L\to\pi^0\nu\bar{\nu}$ decay, 
   these two rare kaon decays become important parts of the search for an 
   understanding of physics beyond the SM. In both channels the decay amplitudes are dominated by SD contributions.
   For $K_L$ decay, the LD contribution can be safely neglected. For $K^+$ decay, the LD effects are expected to be of a few percent, assuming that QCD perturbation theory is accurate at the charm scale.   Although possibly small, this long-distance correction is now the dominate source of theoretical uncertainty in the SM prediction 
   for the $K^+\to\pi^+\nu\bar{\nu}$ branching ratio. It is therefore timely for lattice QCD to provide 
   the LD contribution to $K^+\to\pi^+\nu\bar{\nu}$ with controlled uncertainty.

   In this paper we present a method in which lattice QCD can be used to compute the LD contribution to the $K\to\pi\nu\bar{\nu}$ decay amplitude.  As explained in the body of this paper, the calculation requires the computation of  non-standard correlation functions, the control of SD singularities, the subtraction of unphysical, exponentially growing contributions as the range of the integration over the time separation of the two weak operators is increased and control of finite-volume effects. The principal aim of this paper is to demonstrate that all these challenges can be overcome.
   The computation of the
   $W$-$W$ and $Z$-exchange diagrams is discussed in Sect.~\ref{sec:methodology}. 
   Because of the non-local neutrino structure in the $W$-$W$ diagrams,
   we must include the neutrino and anti-neutrino explicitly in the final state.  In addition,
   we also need to include a lepton propagator in the lattice calculation. In Sec.~\ref{sec:WW} and Appendix~\ref{sec:WW_scalar_amp}, 
   we show in some detail on how to
   deal with the complicated, non-local neutrino structure. The procedure needed to remove the exponentially growing contamination that accompanies the proposed Euclidean-space lattice methods,  from the $W$-$W$ diagrams
   is discussed in detail in Appendix~\ref{sec:ground_state_WW}.  For both the $W$-$W$ and $Z$-exchange diagrams, 
   the lattice amplitudes will have 
   ultra-violet, logarithmic divergences, which are cut off by the  lattice spacing. 

We discuss in Sec.~\ref{sec:SD} 
   on how to perform the necessary SD correction using an extension of the Rome-Southampton method. Power-law, FV corrections are discussed in Sec.~\ref{sec:FV} with an emphasis
   on their natural phase-space suppression. For the $W$-$W$ diagram, to evaluate the FV correction one needs to compute the 
   $K^+\to\pi^0$ and $\pi^0\to\pi^+$ transition amplitudes. For the $Z$-exchange diagram, the FV effects 
   are suppressed significantly either by limited phase-space 
   or by the $\Delta I=1/2$ rule. Only after reaching sub-percent precision, might one need to include the FV 
   corrections from the $\pi^+\pi^0$ intermediate state. As we show above, 
   it is straightforward to extend the FV correction formula needed for the $K_L$-$K_S$ mass 
   difference~\cite{Christ:2015pwa} to the present case of rare kaon decay.

   Using the methods developed in Ref.~\cite{Christ:2015aha} and this paper, it is now possible to undertake  exploratory numerical 
   calculations of the LD contributions to both the $K\to\pi\ell^+\ell^-$~\cite{Christ:2016awg} and $K\to\pi\nu\bar{\nu}$~\cite{Christ:2016xxx} decay amplitudes. This is important not only for providing needed LD information to the SM prediction for these rare kaon decays but also for extending our ability to compute a wider array of  important physical
   observables using the methods of lattice QCD.

    \begin{acknowledgments} We gratefully acknowledge many helpful discussions with our colleagues from the RBC-UKQCD collaboration.
    C.S. warmly thanks Augusto Cecucci and Cristina Lazzeroni for teaching him about the capabilities and schedule of the NA62 experiment.  N.C. and X.F. were supported in part by U.S. DOE grant \#De-SC0011941 while A.P and C.T.S. were supported in part by UK STFC Grant ST/L000296/1 and A.P. additionally by ST/L000458/1.
    \end{acknowledgments}

\appendix
\section{Connection between Euclidean and Minkowski amplitudes}
\label{sec:M-E}
As the methods of lattice QCD are applied to more complex quantities the issue of the formalism used to present the results becomes more important.   The targets of a lattice QCD calculation, such as that presented here, are physical amplitudes which can be compared with other experimental and theoretical work and would naturally be presented as Minkowski space quantities in which the operators involved have a conventional, physical time dependence and Lorentz symmetry is manifest.  However, a lattice QCD calculation requires the introduction of an unphysical, Euclidean time and a resulting formalism that has a Euclidean $O(4)$ symmetry.  

Both descriptions of relativistic quantum field theory can be viewed as based on the same Schr\"odinger quantum mechanics, described by the same quantum mechanical Hilbert space and the same QCD Hamiltonian.  This makes it possible to establish that certain quantities computed using Euclidean-space lattice methods are identical to those of physical interest described using Minkowski time dependence.  However, a given physical quantity will often be expressed using different conventions depending on which approach is adopted, creating a dilemma for a paper such as this.  While we would like to present results in a standard notation immediately accessible to those familiar with Minkowski field theory, we also wish to present a record of our calculation without a translation into a second formalism.

As a compromise we have presented the details of our method in the $O(4)$-invariant, Euclidean formalism used for the calculation but also give important formulae in a conventional, Minkowski language.  In this Appendix we discuss the relation between these two descriptions so that the reader can interpret our Euclidean-space formulae in terms of Minkowski quantities.  This appendix is divided into two sections.  The first, included for completeness, recalls the standard relationship between time-independent quantities computed using Euclidean and Minkowski conventions.  In the second section we specialize these considerations to the quantities computed in this paper and provide the Minkowski-space definitions of those quantities.

\subsection{General considerations}

Starting with the same Schr\"odinger operator $O_S$ the Minkowski and Euclidean approaches define two different time-dependent generalizations:
\begin{eqnarray}
O_M(t) &=& e^{iHt}O_S e^{-iHt} 
\label{eq:M} \\
O_E(x_0) &=& e^{Hx_0}O_S e^{-Hx_0} 
\label{eq:E}
\end{eqnarray}
where $H$ is the QCD Hamiltonian, the subscripts $M$ and $E$ identify Minkowski and Euclidean operators and we use different variables  $t$ and $x_0$ to represent Minkowski and Euclidean time.  

When expressed as a Feynman path integral the time-ordered product of $N$ time-dependent operators,
\begin{equation}
\bigr\langle 0\bigr|T\bigl(O_{X_1}(x_1) O_{X_2}(x_2) \ldots O_{X_N}(x_N)\bigr)\bigr|0\bigr\rangle
\label{eq:time_ordered}
\end{equation}
can be written as manifestly Lorentz- or  $O(4)$-invariant quantities when $X=M$ or $E$, respectively.  While such Green's functions can be viewed as a single analytic function of the space time coordinates $\{x_1, x_2, \ldots,x_N\}$, for numerical work the possibility of performing an analytic continuation is rarely of direct value.  Instead special constructions are employed for the Euclidean-space lattice QCD calculation to extract quantities with direct physical meaning.  Masses of low-lying states can be obtained from the exponential dependence on the time separation of the operators appearing in the Euclidean time-ordered product in Eq.~\eqref{eq:time_ordered} for the case $N=2$.  Likewise the matrix element of a Schr\"odinger operator $O_S$ between physical, energy eigenstates can be obtained from the time-ordered product in Eq.~\eqref{eq:time_ordered} for the case $N=3$ where large time separations are used to project onto the desired energy eigenstates.  For the more complex, bilocal operators considered in this paper, more effort must be expended to extract quantities of physical interest from time integrals of Euclidean time ordered products of the sort shown in Eq.~\eqref{eq:time_ordered} for the case $N=4$.

However, we do not conventionally work with the underlying Schr\"odinger operators, which typically contain conjugate field variables $\pi(x)$ and the Dirac creation operators $\psi^\dagger(x)$.  Instead, these non-covariant, Hamiltonian quantities are replaced by $\partial \phi(x)/\partial x_0$ or $\partial \phi(x)/\partial t$ and $\overline{\psi}(x)$ using conventions that differ between the Minkowski- and Euclidean-space formalisms.  While the treatment of spatial variables should be the same in these two approaches, our use of a $(1,-1,-1,-1)$ signature for the Minkowski space metric introduces an additional minus sign discrepancy with Euclidean quantities which use a metric with the $(1,1,1,1)$ signature.  (For Minkowski-space, we follow the  conventions of Peskin and Schroeder~\cite{Peskin:1995ev} and view the combination $(t,x^1,x^2,x^3)$ as a raised-index, Minkowski-space vector.)  

For a scalar operator $\phi_X(0,\vec x)$ at $x_0=t=0$ there is no difference between the Euclidean and Minkowski versions which implies that $\nabla_i \phi_M(0,\vec x) = \nabla_i \phi_E(0,\vec x)$, $1 \le i \le 3$.   However, as implied by Eqs.~\eqref{eq:M} and \eqref{eq:E}, their time derivatives will differ:
\begin{equation}
\left.\frac{\partial\phi_M(t,\vec x)}{\partial t}\right|_{t=0}
                 = \left.i\frac{\partial\phi_E(x_0,\vec x)}{\partial x_0}\right|_{x_0=0}
\end{equation}
For example, if $\phi_i(x)$ is the i$^{th}$ component of the three-component, isovector pion field operator we can compare the Minkowski and Euclidean space expressions:
\begin{eqnarray}
\frac{\partial}{\partial x_M^\mu}\langle0| \phi_{M,i}(t, \vec x)|\pi(j,\vec p)\rangle
           &=& -i\bigl(\sqrt{m_\pi^2+\vec p\,^2}, -\vec p\bigr)Z_\pi \delta_{ij} e^{-i p_M \cdot x_M} \\
\frac{\partial}{\partial x_E^\mu} \langle0|\phi_{E,i}(x_0, \vec x)|\pi(j,\vec p)\rangle
           &=&  \bigl(-\sqrt{m_\pi^2+\vec p\,^2}, i\vec p\bigr)Z_\pi \delta_{ij} e^{i p_E \cdot x_E}.
\end{eqnarray}
where the state $|\pi(j,\vec p)\rangle$ describes a physical pion with isospin index $j$ and three momentum $\vec p$, $m_\pi$ is the pion mass and $Z_\pi$ is a normalization factor appropriate for the pion interpolating operator $\phi_i(x)$.  The Minkowski and Euclidean four-momentum assocated with this on-shell, pion state are given by:
 \begin{eqnarray}
p_M^\mu &=& (\sqrt{m_\pi^2 +\vec p\,^2}, \vec p)
\label{eq:M_on-shell} \\
p_E^\mu &=& (i\sqrt{m_\pi^2+\vec p\,^2}, \vec p).
\label{eq:E_on-shell}
\end{eqnarray}

For fermions a similar translation between $\overline{\psi}_M$ and $\overline{\psi}_E$ is needed.  Recall that in Dirac's original notation uses the Hamiltonian operator
\begin{equation}
H_D = \int d^3 x \; \psi^\dagger_S(\vec x)
                             \Bigr(\vec\alpha\cdot (-i\vec\nabla) + \beta m\Bigr)\psi_S(\vec x)
\label{eq:Dirac}
\end{equation}
where the Schr\"odinger operators $\psi(\vec x)$ and its hermitian conjugate $\psi^\dagger(\vec x)$ are time-independent and obey the usual anti-commutation relation, $\{\psi^\dagger(\vec x),\psi(\vec y)\} = \delta^3(\vec x - \vec y)$ while the four, $4\times4$, hermitian, Dirac matrices $\vec \alpha$ and $\beta$ are anti-commuting and each have a square which is the identity matrix.  

If the time evolution operator for the Hamiltonian $H_D$ in Eq.~\eqref{eq:Dirac} is written as a Grassmann path integral following the usual textbook derivation~\cite{ZinnJustin:2002ru}, one finds
\begin{eqnarray}
\mathrm{Tr}\left\{ T\Bigl[e^{-H_D T}\psi(y) \psi^\ddagger(z)\Bigr]\right\} && 
\label{eq:Grassmann}  \\
&&\hskip -1.0 in  = \int d[\overline{\chi}] d[\psi]
 \exp\Biggl\{-\int d^3 x\int_0^T d x_0 \;\overline{\chi}\Bigl(\frac{\partial}{\partial x_0} + \vec\alpha\cdot(-i \vec\nabla) +\beta m\Bigr)\psi\Biggr\} \psi(y) \overline{\chi}(z),
\nonumber
\end{eqnarray}
where to be concrete we consider the case of a two-point function.  The fermion field operators $\psi$ and $\psi^\ddagger$\footnote{We have used the operator $\psi^\ddagger$ to represent the Euclidean time evolution of the operator $\psi^\dagger$ which must be distinguished from the hermitian conjugate of the Euclidean time evolution of the operator $\psi$.} have been replaced by the Grassmann integration variables $\psi(x)$ and $\overline{\chi}(x)$ and the Minkowski case can be obtained by inserting a factor of $i$ in front of the Hamiltonian on the left and right-hand sides of Eq.~\eqref{eq:Grassmann} and replacing the Euclidean time variable $x_0$ by $t$.  In each case, we redefine auxiliary Grassmann field $\overline{\chi}$ to give the mass term its standard form and introduce $\gamma$ matrices chosen to make the underlying Lorentz or $O(4)$ symmetry manifest.  

This can be accomplished by the following choices:
\begin{eqnarray}
\overline{\psi_M} &=& \overline{\chi}\beta, 
  \quad \gamma_M^0 = \beta,
  \quad \vec\gamma_M = \beta\vec\alpha \\
\overline{\psi_E} &=& \overline{\chi}\beta, 
  \quad \gamma_E^0 = \beta,
  \quad \vec\gamma_E = -i\beta\vec\alpha.
\end{eqnarray}
With these conventions Eq.~\eqref{eq:Grassmann} and its Minkowski counterpart become
\begin{eqnarray}
\mathrm{Tr}\left\{ T\Bigl[e^{-iH_D T}\psi_M(y) \psi^\ddagger_M(z)\Bigr]\right\}&&  \\
 && \hskip -1.5 in =  \int d[\overline{\psi}_M] d[\psi_M]
 \exp\Biggl\{i\int d^3 x\int_0^T d x_0 \;\overline{\psi}_M\Bigl(\gamma_M^\mu \frac{\partial}{\partial x^\mu} - m\Bigr)\psi_M\Biggr\} 
        \psi_M(y) \overline{\psi}_M(z)\beta
 \nonumber \\
\mathrm{Tr}\left\{ T\Bigl[e^{-H_D T}\psi_E(y) \psi^\ddagger_E(z)\Bigr]\right\}&& \\ 
&& \hskip -1.5 in =  \int d[\overline{\psi}_E] d[\psi_E]
 \exp\Biggl\{-\int d^3 x\int_0^T d x_0
   \;\overline{\psi}_E\Bigl(\gamma_E^\mu \frac{\partial}{\partial x^\mu}+m \Bigr)\psi_E\Biggr\} 
        \psi_E(y) \overline{\psi}_E(z)\beta
\nonumber
\end{eqnarray}

Thus, the relation between fermionic quantities expressed in the Euclidean and Minkowski formalisms is also straight-forward.  When evaluated at zero time, the Grassmann spinor variables $\overline{\psi}_M(0,\vec x)\beta$ and $\overline{\psi}_E(0,\vec x)\beta$ both correspond to the Schr\"odinger operator $\psi_S^\dagger(\vec x)$, the same relation which connects $\psi(0,\vec x)_M$ and $\psi(0,\vec x)_E$ and $\psi_S(\vec x)$. The Euclidean and Minkowski $\gamma$ matrices are related by
\begin{equation}
\gamma^0_E=\gamma^0_M, \quad \gamma^i_E=-i\gamma^i_M.
\label{eq:E-M_gamma}
\end{equation}
With these rules we can easily relate operators which are expressed in these two formalisms as will be done below.

First we examine the isovector current, normalized so that the integral of the time component generates isospin transformations.  In the case of a scalar field we have:
\begin{eqnarray}
\bigl(\vec V_M^0,\vec V_M^i\bigr) &=& \frac{1}{i}\Biggl(\frac{\partial}{\partial t}\vec \phi \times \vec \phi, - \frac{\partial}{\partial x^i} \vec \phi \times \vec \phi\Biggr) \\
\bigl(\vec V_E^0,\vec V_E^i\bigr) &=& \Biggl(\frac{\partial}{\partial x_0}\vec \phi \times \vec \phi, \frac{\partial}{\partial x^i} \vec \phi \times \vec \phi\Biggr)
\end{eqnarray}
where the explicit vector arrows represents the isospin degree freedom.  Thus, the Minkowski and Euclidean current operators are related by
\begin{equation}
\vec V_M^0 = \vec V_E^0,\quad \vec V_M^i = i\vec V_E^i.
\label{eq:E-M_current}
\end{equation}
We find the same relation if we consider the vector current constructed from fermions which are assumed to form an isodoublet:
\begin{equation}
\vec V_X^\mu = \overline{\psi}_X\gamma^\mu_X\vec \tau \psi_X
\label{eq:E-M_current}
\end{equation}
for $X=M$ or $E$ and $\vec\tau$ is a vector formed from the standard Pauli matrices $\tau^i$.  That the relation in Eq.~\eqref{eq:E-M_current} holds in this case as well as can be deduced from the relation between the Euclidean and Minkowski gamma matrices given in Eq.~\eqref{eq:E-M_gamma}.  The same relation will connect the Euclidean and Minkowski axial currents since in both cases we use the same $\gamma^5$ Dirac matrix: $\gamma_5 = i\gamma_M^0\gamma_M^1\gamma_M^2\gamma_M^3$.

Finally we consider the relation between the four-fermion operators expressed in Euclidean or Minkowski notation.  This is particularly simple because these have the form $\overline{\psi}_X \Gamma_X^i\psi_X \overline{\psi}_X \Gamma_X^j \psi C_X^{ij}$ where $X=M$ or $E$, the $\Gamma_X$ are combinations of spinor and flavor matrices and the coefficients $C_X^{ij}$ are chosen so that the resulting operator is a scalar under the proper Lorentz group or $O(4)$.  Such a quantity is the same for either Minkowski or Euclidean conventions because the four-vector indices of all internal gamma matrices must be contracted in pairs of the form $\gamma_X^\mu\cdots{\gamma_X}_\mu$, a combination which is the same for $X=E$ or $X=M$.

\subsection{Minkowski-space definitions}

Using the above results we will now discuss some specific matrix elements and invariant functions used in this paper and the form in which they appear in both the Euclidean and Minkowski space formalisms.  We use the usual relativistic normalization for single-particle energy eigenstates $|\vec p\rangle$  with mass $m$ carrying momentum $\vec p$
\begin{equation}
\langle \vec p\,'|\vec p\rangle = 2\sqrt{\vec p^2 + m^2}(2\pi)^3 \delta^3(\vec p\,' -\vec p). 
\end{equation}
For spin-1/2 particles, we will introduce the usual positive and negative energy spinor eigenstates of the free Dirac Hamiltonian $\vec \alpha\cdot\vec p + \beta m$,  $u(\vec p,s)$ and $v(-\vec p,s)$ corresponding to particle and anti-particle states with spin $s$,  normalized so that the projection operators $P_\pm$ onto states of both spins with positive or negative energy take the form:
\begin{eqnarray}
P_+ &=& \sum_{s=\pm\frac{1}{2}} u(\vec p)u(\vec p)^\dagger
        = \vec\alpha\cdot\vec p + \beta m+E
        = \left(\gamma_M^\mu {p_M}_\mu + m \right)\beta
        = \left(-i\gamma_E^\mu {p_E}_\mu + m \right)\beta \\
P_-  &=& \sum_{s=\pm\frac{1}{2}} v(\vec p)v(\vec p)^\dagger
        = \vec\alpha\cdot\vec p - \beta m +E
        = \left(\gamma_M^\mu {p_M}_\mu - m \right)\beta 
        = \left(-i\gamma_E^\mu {p_E}_\mu - m \right)\beta 
\end{eqnarray}
where $E=\sqrt{\vec p^2 + m^2}$. These same two $4\times4$ projection operators can be used to compute polarization sums from products of matrix elements that were computed using either Minkowski or Euclidean conventions.   Of course, the covariant Euclidean and Minkowski expressions in these equations require that the appropriate on-shell momentum given in Eqs.~\eqref{eq:E_on-shell} and \eqref{eq:M_on-shell} be used. 

The most familiar matrix element to describe is that defining the pseudoscalar decay constant $f_\pi$ for which we can write both Euclidean- and Minkowski-space expressions as dictated by Eq.~\eqref{eq:E-M_current}:
\begin{eqnarray}
\langle0|[\bar{d}\gamma^\mu\gamma_5u]_M(x_M)|\pi^+(\vec{p})\rangle
      &=& i{p_M}^\mu f_\pi e^{-i(E_\pi t -\vec{p}\cdot\vec{x})} \\
\langle0|[\bar{d}\gamma^\mu\gamma_5u]_E(x_E)|\pi^+(\vec{p})\rangle
      &=& {p_E}^\mu f_\pi e^{-E_\pi x_0+i\vec{p}\cdot\vec{x}}.
\end{eqnarray}

A second example is the matrix element of the vector current between charged kaon and pion states:
\begin{eqnarray}
\langle\pi^+(\vec p_\pi)|\bar{s}\gamma_M^\mu d(0)|K^+(\vec p_K)\rangle 
  &=& -\left(f_+(q_M^2) (p_K+p_\pi)_M^\mu +f_-(q_M^2)(p_K-p_\pi)_M^\mu\right) \\
\langle\pi^+(\vec p_\pi)|\bar{s}\gamma_E^\mu d(0)|K^+(\vec p_K)\rangle
  &=& i\left(f_+(-q_E^2)(p_K+p_\pi)_E^\mu+f_-(-q_E^2)(p_K-p_\pi)_E^\mu\right).
\end{eqnarray}
Here the minus signs in the arguments of $f_\pm(q^2)$ in the Euclidean expression ensure that precisely the same form factors enter both expressions, compensating for the different signs in the inner product that result when equivalent momenta are used in our Euclidean and Minkowski conventions.

Finally we examine the matrix elements of the bilinear operators which are the primary topic of this paper.  In such four-point correlation functions, the individual four-fermion operators $\{O,O'\}=\{O_{q\ell}^{\Delta S=1}$, $O_{q\ell}^{\Delta S=0}\}$ for the $W$-$W$ diagram and \{$O_q^W$, $O_\ell^Z$\} for the $Z$-exchange diagram are all scalar operators and hence the same in both Euclidean and Minkowski conventions.  In Ref.~\cite{Buchalla:1993wq}, the Minkowski expression for the bilocal operator product has been defined as
\ba
\mathcal{B}_M=i\int d^4x_M\,T[O_M(x_M)\,O_M'(0)]-\{u\to c\}.
\ea

The physical, Minkowski-space transition amplitude $A_M=\langle f|{\mathcal B}_M|i\rangle$ with initial state $|i\rangle$ and final state $|f\rangle$ can be written as
\ba
A_M&=&i\int_0^\infty d t \sum_n\langle f|O_M|n\rangle\langle n|O_M'|i\rangle e^{i(E_f-E_n)t}
\nn\\
&+&i\int_{-\infty}^0 d t \sum_k\langle f|O_M'|m\rangle\langle m|O_M|i\rangle e^{i(E_m-E_i)t}-\{u\to c\}
\nn\\
&=&-\sum_n\frac{\langle f|O_M|n\rangle\langle n|O_M'|i\rangle}{E_f-E_n+i\varepsilon}
+\sum_m\frac{\langle f|O_M'|m\rangle\langle m|O_M|i\rangle}{E_m-E_i-i\varepsilon}-\{u\to c\}
\ea
The corresponding Euclidean expression is given by
\ba
\mathcal{B}_E&=&\int d^4x_E\,T[O_E(x_E)\,O_E'(0)]-\{u\to c\}
\nn\\
&=&\int_{-T_a}^{T_b} d x_0 \int d^3 x \,T[O_E(x_E)\,O_E'(0)]-\{u\to c\}.
\ea
The transition amplitude $A_E=\langle f|{\mathcal B}_E|i\rangle$ is then given by
\ba
A_E&=&
-\sum_n\frac{\langle f|O_E|n\rangle\langle n|O_E'|i\rangle}{E_f-E_n}\left(1-e^{(E_f-E_n)T_b}\right)
\nn\\
&&+\sum_m\frac{\langle f|O_E'|m\rangle\langle m|O_E|i\rangle}{E_m-E_i}\left(1-e^{(E_i-E_m)T_a}\right)-\{u\to c\}
\ea
The equality of the matrix elements $\langle f|O_E|n\rangle$ and $\langle f|O_M|n\rangle$ then guarantees that $A_E$ is equal to $A_M$ once we have removed the exponentially growing contamination in $A_E$.

\section{Mesonic and leptonic states}
\label{sec:states}
    The mesonic states used in this paper are defined as the lowest energy component of the state that results from applying the following combinations of quark and anti-quark operators to the QCD vacuum state. (Here we are only concerned with the flavor and sign conventions so detailed questions of the spatial structure of the combination of quark and anti-quark operators are not addressed.)
    \begin{eqnarray} 
    &&|\pi^+\rangle=i\bar{u}\gamma_5d|0\rangle,
    \quad |\pi^-\rangle=-i\bar{d}\gamma_5u|0\rangle,
    \quad |\pi^0\rangle=\frac{i}{\sqrt{2}}(\bar{u}\gamma_5u-\bar{d}\gamma_5d)|0\rangle
    \nn\\
    &&|K^+\rangle=i\bar{u}\gamma_5s|0\rangle,
    \quad |K^-\rangle=-i\bar{s}\gamma_5u|0\rangle,
    \quad |K^0\rangle=i\bar{d}\gamma_5s|0\rangle,
    \quad |\overline{K}^0\rangle=-i\bar{s}\gamma_5d|0\rangle.
\label{eq:mesons}
    \nn\\
    \end{eqnarray} 
    In an analogous fashion, leptonic states can be annihilated by the corresponding leptonic field operators, leaving the usual Dirac plane-wave spinors
    \begin{eqnarray}
    && \nu(x)|\nu(p_\nu)\rangle=u(p_\nu)e^{ip_\nu x}|0\rangle,
    \quad \bar{\nu}(x)|\bar{\nu}(p_{\bar{\nu}})\rangle=\bar{v}(p_{\bar{\nu}})e^{ip_{\bar{\nu}}x}|0\rangle
    \nn\\
    && \ell(x)|\ell(p_\ell)\rangle=u(p_\ell)e^{ip_\ell x}|0\rangle,
    \quad \bar{\ell}(x)|\bar{\ell}(p_{\bar{\ell}})\rangle=\bar{v}(p_{\bar{\ell}})e^{ip_{\bar{\ell}}x}|0\rangle,
\label{eq:leptons}
    \end{eqnarray}
    where the spinors $u(p)$ and $v(p)$ are the conventional positive- and negative-energy eigenvectors of the Dirac Hamiltonian introduced in Appendix~\ref{sec:M-E}.  Note the spinor $u$ in Eq.~\eqref{eq:leptons} should not be confused with the up quark operator appearing in Eq.~\eqref{eq:mesons}.
    For simplicity we have not shown the spin index. 

\section{Extraction of the scalar amplitude from {\boldmath$W$-$W$} diagrams}
\label{sec:WW_scalar_amp}
We write the integrand in the bilocal matrix element $T_{WW}$ defined in Eq.\,(\ref{eq:TWW}) in terms of two factors:
   \begin{equation}
   \label{eq:H_and_L}
   T_{WW}=\int d^4x\,H_{\alpha\beta}(x)\,\left[\bar{u}(p_\nu)\Gamma_{\alpha\beta}(x){v}(p_{\bar{\nu}})\right]\,.
   \end{equation}
   The hadronic factor $H_{\alpha\beta}(x)$ and the leptonic factor  
   $\bar{u}(p_\nu)\Gamma_{\alpha\beta}(x){v}(p_{\bar{\nu}})$ are defined by
   \begin{eqnarray}
   H_{\alpha\beta}(x)&=&Z_V^2\langle\pi^+(p_\pi)|T[\bar{s}\gamma_\alpha
   (1-\gamma_5)u(x)\,\bar{u}\gamma_\beta(1-\gamma_5)d(0)]|K^+(p_K)\rangle-\{u\to c\}
   \nn\\
   \Gamma_{\alpha\beta}(x)&=&\gamma_\alpha
   (1-\gamma_5)S_\ell(x,0)\gamma_\beta(1-\gamma_5)e^{ip_\nu x}.
   \end{eqnarray}
   Here $S_{\ell}(x,0)=\int\frac{d^4q}{(2\pi)^4}\frac{-i{\slashed q}+m_\ell}{q^2+m_\ell^2}e^{iqx}$ is a free Euclidean lepton propagator.

   The left-handed nature of neutrinos allows us to write $T_{WW}$ in the form
   \begin{equation}\label{eq:TWW2}
   T_{WW}=T_\mu\,\bar{u}(p_\nu)\gamma_\mu(1-\gamma_5){v}(p_{\bar{\nu}})\,,
   \end{equation}
   where 
   with three independent momenta $p_K$, $p_\nu$ and $p_{\bar{\nu}}$, $T_\mu$ can be written as
   \begin{equation}\label{eq:T1234}
   T_\mu={p_K}_\mu G_1+{p_\nu}_\mu G_2+{p_{\bar{\nu}}}_\mu G_3+\varepsilon_{\mu\alpha\beta\rho}{p_K}_\alpha
   {p_\nu}_\beta {p_{\bar{\nu}}}_\rho G_4.
   \end{equation}
   Neglecting the masses of the neutrinos, the terms proportional to ${p_\nu}_\mu$ and ${p_{\bar{\nu}}}_\mu$ vanish because of 
the Dirac equation obeyed by the neutrino wave function.
   
   We now consider the term proportional to $G_4$ in Eq.~\eqref{eq:T1234}.  Using the identity
   $\gamma_\alpha\gamma_\beta\gamma_\rho=\delta_{\alpha\beta}\gamma_\rho+\delta_{\beta\rho}\gamma_\alpha
   -\delta_{\alpha\rho}\gamma_\beta+\varepsilon_{\mu\alpha\beta\rho}\gamma_\mu\gamma_5$
   we can write
   \begin{equation}
   \label{eq:gamma_matrix}
   \varepsilon_{\mu\alpha\beta\rho}{p_K}_\alpha{p_\nu}_\beta{p_{\bar{\nu}}}_\rho
   \gamma_\mu(1-\gamma_5)
   =-\left[{\slashed p}_K{\slashed p}_\nu{\slashed p}_{\bar{\nu}}-(p_K\cdot p_\nu){\slashed p}_{\bar{\nu}}
   -(p_\nu\cdot p_{\bar{\nu}}){\slashed p}_K+(p_K\cdot p_{\bar{\nu}}){\slashed p}_\nu\right](1-\gamma_5).
   \end{equation}
   Since the right-hand side of Eq.\,(\ref{eq:gamma_matrix}) is sandwiched between the neutrino spinors $\bar{u}(p_\nu)$ and $v(p_{\bar{\nu}})$ in Eq.\,(\ref{eq:TWW2}),
   only the third term in Eq.\,(\ref{eq:gamma_matrix})
   survives. Thus,  when $T_\mu$ is combined with the product of neutrino spinors 
   in Eq.\,(\ref{eq:TWW2}), the term proportional to $G_4$ in Eq.\,(\ref{eq:T1234}) is also effectively proportional to $p_{K_\mu}$. Therefore, we can write $T_{WW}$ in terms of a single invariant amplitude $F_{WW}$:
\begin{equation}
   \label{eq:W_amplitude}
   \int d^4x\,H_{\alpha\beta}(x)\,\left[\bar{u}(p_\nu)\Gamma_{\alpha\beta}(x){v}(p_{\bar{\nu}})\right]
   =i\cdot F_{WW}(p_K,p_\nu,p_{\bar{\nu}})\,\left[
   \bar{u}(p_\nu){\slashed p}_K(1-\gamma_5){v}(p_{\bar{\nu}})\right]\,.
   \end{equation}
 
 We now derive an expression for the scalar amplitude 
$F_{WW}(p_K,p_\nu,p_{\bar{\nu}})$.   This might be most naturally done by following
the steps that are taken when evaluating the $K^+\to\pi^+\nu\bar\nu$ decay rate.   Thus,
we multiply both sides of Eq.~\eqref{eq:W_amplitude} by the $2\times2$ spin matrix $\bar{v}{\slashed p}_K(1-\gamma_5) u$ and perform the spin sums in order to project out $F_{WW}$ obtaining
   \begin{equation}\label{eq:FWW2}
   F_{WW}(p_K,p_\nu,p_{\bar{\nu}})=\frac{-i\int d^4x\,H_{\alpha\beta}(x)\,
   \textmd{Tr}[\Gamma_{\alpha\beta}(x){\slashed p}_{\bar{\nu}}
  {\slashed p}_K(1-\gamma_5){\slashed p}_\nu]}{\textmd{Tr}[{\slashed p}_K(1-\gamma_5)
  {\slashed p}_{\bar{\nu}}{\slashed p}_K(1-\gamma_5){\slashed p}_\nu]}.
   \end{equation}
   
   For lattice calculations it is useful to simplify the expression on the right-hand side of Eq.\,(\ref{eq:FWW2}).
   The gamma matrix factor ${\slashed p}_{\bar{\nu}}{\slashed p}_K(1-\gamma_5){\slashed p}_\nu$, which appears in both the traces in the numerator and the 
   denominator,
   can be rewritten in the form
   \begin{eqnarray}
   {\slashed p}_{\bar{\nu}}{\slashed p}_K(1-\gamma_5){\slashed p}_\nu
   =\sum_\mu b_\mu\gamma_\mu(1+\gamma_5),
   \end{eqnarray}
   where the coefficient $b_\mu$ given by
   \begin{eqnarray} 
   b_\mu&=&\frac{1}{4}{\textmd{Tr}}[\gamma_\mu{\slashed p}_{\bar{\nu}}{\slashed p}_K(1-\gamma_5){\slashed p}_\nu]
   \nn\\
   &=&{p_{\bar{\nu}}}_\mu(p_K\cdot p_\nu)+{p_{\nu}}_\mu(p_K\cdot p_{\bar{\nu}})-{p_K}_\mu(p_\nu\cdot p_{\bar{\nu}})
   +\varepsilon_{\mu\alpha\beta\rho}{p_\nu}_\alpha {p_{\bar{\nu}}}_\beta {p_K}_\rho.
   \end{eqnarray}
   This allows us to rewrite $F_{WW}(p_K,p_\nu,p_{\bar{\nu}})$ in the form
   \begin{equation}
   F_{WW}(p_K,p_\nu,p_{\bar{\nu}})=-i\int d^4x\,H_{\alpha\beta}(x)\,
   \sum_\mu c_\mu\textmd{Tr}\left[\Gamma_{\alpha\beta}(x)\gamma_\mu(1+\gamma_5)\right],
   \end{equation}
   where the four-vector $c_\mu$ is given by
   \begin{equation}
   c_\mu=\frac{1}{8}\frac{b_\mu}{b\cdot p_K}\,.
   \end{equation}
   Given the momenta $p_K$, $p_\nu$ and $p_{\bar{\nu}}$, the coefficients $c_\mu$ can readily be evaluated so we need to compute only the four integrals $\int d^4x\,H_{\alpha\beta}(x)\,
   \textmd{Tr}[\Gamma_{\alpha\beta}(x)\gamma_\mu(1+\gamma_5)]$ for $\mu=$0, 1, 2 and 3.

   In a lattice calculation, the hadronic matrix element $H_{\alpha\beta}(x)$ can be calculated by
   evaluating a 4-point correlation function. The leptonic propagator $S_\ell(x,0)$ in $\Gamma_{\alpha\beta}(x)$
   can be implemented using a free-field lattice fermion formulation, e.g.
   domain wall or overlap fermion. Following the steps described above one
   can determine the scalar amplitude $F_{WW}(p_K,p_\nu,p_{\bar{\nu}})$.

\section{Low-lying intermediate states for {\boldmath$W$-$W$} diagrams}
\label{sec:ground_state_WW}
   As indicated in Sec.~\ref{sec:exp_grow_contamination}, if the energy of a given intermediate state is 
   smaller than the energy of initial/final state,
   then in Euclidean space-time, the non-local matrix element 
   $\int dt\,\langle\pi^+\nu\bar{\nu}|T[O^{\Delta S=1}_{u\ell}(t)O^{\Delta S=0}_{u\ell}(0)]|K^+\rangle$
   will include an exponentially growing contamination.
   Here we study what we expect will be the largest exponentially growing contamination from the low-lying intermediate states.

   For $t\ll0$, the non-local matrix element is 
   dominated by the intermediate ground state $|\bar{\ell}\nu\rangle$.  Its time dependence can be written as 
   \begin{eqnarray}
   &&\langle\pi^+\nu\bar{\nu}|O_{u\ell}^{\Delta S=0}(0)|\bar{\ell}\nu\rangle\frac{1}{2E_{\bar{\ell}}}\frac{1}{2E_\nu}
   \langle\bar{\ell}\nu|O_{u\ell}^{\Delta S=1}(t)|K^+\rangle
   \nn\\
   &=&Z_V\langle\pi^+|\bar{u}\gamma_\mu(1-\gamma_5)d(0)|0\rangle\,Z_V\langle0|\bar{s}\gamma_\nu(1-\gamma_5)u(0)|K^+\rangle
   \nn\\
   &&\times\bar{u}(p_\nu)\gamma_\nu(1-\gamma_5)\frac{i{\slashed p}_{\bar{\ell}}+m_{\bar{\ell}}}{2E_{\bar{\ell}}}
   \gamma_\mu(1-\gamma_5)v(p_{\bar{\nu}})\cdot e^{(E_{\bar{\ell}}+E_\nu-E_K)t}
   \nn\\
   &=&-2f_Kf_\pi\bar{u}(p_\nu){\slashed p}_K\frac{i{\slashed p}_{\bar{\ell}}}{2E_{\bar{\ell}}}
   {\slashed p}_\pi(1-\gamma_5)v(p_{\bar{\nu}})\cdot e^{(E_{\bar{\ell}}+E_\nu-E_K)t}
   \nn\\
   &\equiv& c_{t<0}\cdot e^{(E_{\bar{\ell}}+E_\nu-E_K)t},
   \label{eq:ground_state_tll0}
   \end{eqnarray}
   where $f_K$ and $f_\pi$ are the kaon and pion decay constants.
   Here we have used the definition $Z_V\langle0|\bar{s}\gamma_\mu\gamma_5u(0)|K^+\rangle={p_K}_\mu f_K$ and
   $Z_V\langle\pi^+|\bar{u}\gamma_\mu\gamma_5d(0)|0\rangle=-{p_\pi}_\mu f_\pi$. The 4-momenta for initial-, intermediate- and
   final-state particles are given by
   \begin{eqnarray}
   p_i=(iE_i,\vec{p}_i),\quad E_i=\sqrt{m_i^2+\vec{p}\,^2_i},\quad i=K,\pi,\nu,\bar{\nu},\bar{\ell}.
   \end{eqnarray}
   Three-momenta conservation requires $\vec{p}_{\bar{\ell}}=\vec{p}_K-\vec{p}_\nu=\vec{p}_\pi+\vec{p}_{\bar{\nu}}$.  (See Fig.~\ref{fig:Wbox}.)

   For $t\gg0$, due to the exchange of the operators $O^{\Delta S=1}_{u\ell}$ and $O^{\Delta S=0}_{u\ell}$, the leptonic part of the
   intermediate state is now given by $\ell\bar{\nu}$. To guarantee the flavor and charge conservation, the hadronic part must be a
   strange state with electric charge $Q_e=+2$. In this case, the lowest energy intermediate state is given by $|K^+\pi^+\ell\bar{\nu}\rangle$.  This four-particle state has an energy larger than that of the kaon and hence will not contribute a growing exponential term.  Note that for this intermediate state, only the 3-momentum of $\bar{\nu}$ is fixed. For the purposes of this analytic treatment we will include the special case in which this intermediate state contains a $K^+$ and $\pi^+$ which do not scatter and carry the same 3-momenta as those of the initial-state kaon and final-state pion respectively.  (Examining this case allow us to show how the non-scattering part of the $K^+\pi^+$ intermediate state contributes to give the usual covariant charged lepton propagator when the two time orderings are combined.)  Including this component of the intermediate $K^+$-$\pi^+$, we have
   \begin{eqnarray}
   \label{eq:ground_state_tgg0}
   &&\langle\pi^+\nu\bar{\nu}|O_{u\ell}^{\Delta S=1}(t)|K^+\pi^+\ell\bar{\nu}\rangle
   \frac{1}{2E_K}\frac{1}{2E_\pi}\frac{1}{2E_\ell}\frac{1}{2E_{\bar{\nu}}}
   \langle K^+\pi^+\ell\bar{\nu}|O_{u\ell}^{\Delta S=0}(0)|K^+\rangle
   \nn\\
   &=&Z_V\langle0|\bar{s}\gamma_\mu(1-\gamma_5)u(0)|K^+\rangle\,Z_V\langle\pi^+|\bar{u}\gamma_\nu(1-\gamma_5)d(0)|0\rangle
   \nn\\  
   &&\times\bar{u}(p_\nu)\gamma_\mu(1-\gamma_5)\frac{-i{\slashed p}_{\ell}+m_\ell}{2E_\ell}
   \gamma_\nu(1-\gamma_5)v(p_{\bar{\nu}})\cdot e^{(E_\nu-E_\ell-E_K)t}
   \nn\\
   &=&-2f_Kf_\pi\bar{u}(p_\nu){\slashed p}_K\frac{-i{\slashed p}_{\ell}}{2E_\ell}
   {\slashed p}_\pi(1-\gamma_5)v(p_{\bar{\nu}})\cdot e^{(E_\nu-E_\ell-E_K)t}
   \nn\\
   &\equiv& c_{t>0}\cdot e^{(E_\nu-E_\ell-E_K)t},
   \end{eqnarray}
   where $p_\ell=(iE_\ell,\vec{p}_\ell)$, $\vec{p}_\ell=-(\vec{p}_K-\vec{p}_\nu)$ and $E_\ell=\sqrt{m_\ell^2+\vec{p}_\ell^2}$.

   Combining the contributions given by Eqs.~(\ref{eq:ground_state_tll0}) and (\ref{eq:ground_state_tgg0}) and
   performing the time integral in a window $[-T_a,T_b]$, we have
   \begin{eqnarray}
   \int_{-T_a}^{0} dt\,c_{t<0}\cdot e^{(E_{\bar{\ell}}+E_\nu-E_K)t}+\int_0^{T_b}dt\,c_{t>0}\cdot e^{(E_\nu-E_\ell-E_K)t} &&
   \nn\\
   && \hskip -3.3 in =\frac{c_{t<0}}{E_{\bar{\ell}}+E_\nu-E_K}\left(1-e^{-(E_{\bar{\ell}}+E_\nu-E_K)T_a}\right)
   -\frac{c_{t>0}}{E_\nu-E_\ell-E_K}\left(1-e^{(E_\nu-E_\ell-E_K)T_b}\right) \hskip 0.3 in
   \\
   \label{eq:term}
  && \hskip -3.3 in = -2f_Kf_\pi\bar{u}(p_\nu){\slashed p}_K\frac{i{\slashed q}}{q^2+m_\ell^2}{\slashed p}_\pi(1-\gamma_5)v(p_{\bar{\nu}})
   \\
   && \hskip -3.0 in -\frac{c_{t<0}}{E_{\bar{\ell}}+E_\nu-E_K}e^{-(E_{\bar{\ell}}+E_\nu-E_K)T_a}
    +\frac{c_{t>0}}{E_\nu-E_\ell-E_K}e^{(E_\nu-E_\ell-E_K)T_b}, \nn
   \end{eqnarray}
   with the 4-momentum $q=p_K-p_\nu$.
   The top term on the right hand side of Eq.~\eqref{eq:term} corresponds to the simplest graph contributing to diagrams of type 1, where the
   process of kaon leptonic decay and (inverse) pion leptonic decay are joined by a lepton propagator. (See Fig.~\ref{fig:Wbox}.)  
   The expression in this term can be further simplified to
   \begin{eqnarray}
   \label{eq:vacuum_saturation}
   (-i)\,f_Kf_\pi\frac{2q^2}{q^2+m_\ell^2}\cdot\bar{u}(p_\nu){\slashed p}_K(1-\gamma_5)v(p_{\bar{\nu}}).
   \end{eqnarray}
   The left term in the lowest line of Eq.~\eqref{eq:term} gives the exponentially growing contamination, which can be
   removed once we evaluate the coefficient $c_{t<0}$ defined in Eq.~(\ref{eq:ground_state_tll0}).
   The right term in the lowest line of Eq.~\eqref{eq:term} vanishes exponentially because $E_\nu<E_\ell+E_K$ and thus requires no special treatment.

   Next, let us look at the second lowest intermediate state.
   For $t\ll0$, it is given by $|\pi^0\bar{\ell}\nu\rangle$ and we have
   \begin{eqnarray}
   \label{eq:type2_ground}
   \int\frac{d^3\vec{p}_{\pi^0}}{(2\pi)^3}
   \langle\pi^+\nu\bar{\nu}|O_{u\ell}^{\Delta S=0}(0)|\pi^0\bar{\ell}\nu\rangle\frac{1}{2E_{\pi^0}}\frac{1}{2E_{\bar{\ell}}}\frac{1}{2E_\nu}
   \langle\pi^0\bar{\ell}\nu|O_{u\ell}^{\Delta S=1}(t)|K^+\rangle &&
   \nn\\
   &&\hskip -3.5 in =\int\frac{d^3\vec{p}_{\pi^0}}{(2\pi)^3}Z_V\langle\pi^+|\bar{u}\gamma_\mu(1-\gamma_5)d(0)|\pi^0\rangle\frac{1}{2E_{\pi^0}}
   Z_V\langle\pi^0|\bar{s}\gamma_\nu(1-\gamma_5)u(0)|K^+\rangle
   \nn\\
   &&\hskip -3.2 in \cdot\bar{u}(p_\nu)\gamma_\nu(1-\gamma_5)\frac{i{\slashed p}_{\bar{\ell}}+m_{\bar{\ell}}}{2E_{\bar{\ell}}}
   \gamma_\mu(1-\gamma_5)v(p_{\bar{\nu}})\cdot e^{(E_{\bar{\ell}}+E_{\pi^0}+E_\nu-E_K)t},
   \end{eqnarray}
   where $\vec{p}_{\pi^0}$ is the 3-momentum of the intermediate neutral pion.  Momentum conservation implies that the anti-lepton carries the 3-momentum
   $\vec{p}_{\bar{\ell}}=\vec{p}_K-\vec{p}_\nu-\vec{p}_{\pi^0}$. Exponentially growing contamination is then associated with
   those intermediate states whose energies satisfy $E_{\bar{\ell}}+E_{\pi^0}+E_\nu<E_K$. This constraint results 
   in a phase-space suppression, which substantially reduces the exponential contamination.

   In a lattice QCD calculation with a finite volume $L^3$, the 3-momentum integral in Eq.~(\ref{eq:type2_ground}) is replaced by a sum
   \begin{eqnarray}
   \int\frac{d^3\vec{p}_{\pi^0}}{(2\pi)^3} \quad \to \quad \frac{1}{L^3}\sum_{\vec{p}_{\pi^0}}.
   \end{eqnarray}
   The scale of a typical lattice momentum is around $2\pi/L\sim2\pi/(4/m_\pi)\sim 220$ MeV. Therefore, in the kaon rest frame, the energies of only a 
   few $|\pi^0\bar{\ell}\nu\rangle$ states will  lie below the energy $E_K=m_K$. For each such state, one can evaluate the hadronic
   matrix elements $\langle\pi^+|\bar{u}\gamma_\mu d(0)|\pi^0\rangle$ and $\langle\pi^0|\bar{s}\gamma_\nu u(0)|K^+\rangle$.
   Thus, the exponentially growing contamination for type 2 diagrams can be removed if observed.

   It is possible that higher energy intermediate states such as $|\pi\pi\bar{\ell}\nu\rangle$ and $|3\pi\bar{\ell}\nu\rangle$ may have energies
   below $E_K$. However, because of an even more suppressed phase space, the exponentially growing contamination from these states will be negligibly
   small. We therefore do not discuss these states in detail.

\bibliography{paper}
\end{document}